\PassOptionsToPackage{unicode}{hyperref}
\PassOptionsToPackage{hyphens}{url}
\PassOptionsToPackage{dvipsnames,svgnames,x11names}{xcolor}
\documentclass[12pt]{article}

\usepackage{amsmath,amssymb}
\usepackage{iftex}
\ifPDFTeX
  \usepackage[T1]{fontenc}
  \usepackage[utf8]{inputenc}
  \usepackage{textcomp} 
\else 
  \usepackage{unicode-math}
  \defaultfontfeatures{Scale=MatchLowercase}
  \defaultfontfeatures[\rmfamily]{Ligatures=TeX,Scale=1}
\fi
\usepackage{lmodern}
\ifPDFTeX\else  
\fi
\IfFileExists{upquote.sty}{\usepackage{upquote}}{}
\IfFileExists{microtype.sty}{
  \usepackage[]{microtype}
  \UseMicrotypeSet[protrusion]{basicmath} 
}{}
\makeatletter
\@ifundefined{KOMAClassName}{
  \IfFileExists{parskip.sty}{%
    \usepackage{parskip}
  }{
    \setlength{\parindent}{0pt}
    \setlength{\parskip}{6pt plus 2pt minus 1pt}}
}{
  \KOMAoptions{parskip=half}}
\makeatother
\usepackage{xcolor}
\makeatletter
\ifx\paragraph\undefined\else
  \let\oldparagraph\paragraph
  \renewcommand{\paragraph}{
    \@ifstar
      \xxxParagraphStar
      \xxxParagraphNoStar
  }
  \newcommand{\xxxParagraphStar}[1]{\oldparagraph*{#1}\mbox{}}
  \newcommand{\xxxParagraphNoStar}[1]{\oldparagraph{#1}\mbox{}}
\fi
\ifx\subparagraph\undefined\else
  \let\oldsubparagraph\subparagraph
  \renewcommand{\subparagraph}{
    \@ifstar
      \xxxSubParagraphStar
      \xxxSubParagraphNoStar
  }
  \newcommand{\xxxSubParagraphStar}[1]{\oldsubparagraph*{#1}\mbox{}}
  \newcommand{\xxxSubParagraphNoStar}[1]{\oldsubparagraph{#1}\mbox{}}
\fi
\makeatother

\usepackage{longtable,booktabs,array}
\usepackage{calc} 
\usepackage{etoolbox}
\makeatletter
\patchcmd\longtable{\par}{\if@noskipsec\mbox{}\fi\par}{}{}
\makeatother
\IfFileExists{footnotehyper.sty}{\usepackage{footnotehyper}}{\usepackage{footnote}}
\makesavenoteenv{longtable}
\usepackage{graphicx}
\makeatletter
\def\maxwidth{\ifdim\Gin@nat@width>\linewidth\linewidth\else\Gin@nat@width\fi}
\def\maxheight{\ifdim\Gin@nat@height>\textheight\textheight\else\Gin@nat@height\fi}
\makeatother
\setkeys{Gin}{width=\maxwidth,height=\maxheight,keepaspectratio}
\makeatletter
\def\fps@figure{htbp}
\makeatother

\makeatletter
\@ifpackageloaded{caption}{}{\usepackage{caption}}
\AtBeginDocument{%
\ifdefined\contentsname
  \renewcommand*\contentsname{Table of contents}
\else
  \newcommand\contentsname{Table of contents}
\fi
\ifdefined\listfigurename
  \renewcommand*\listfigurename{List of Figures}
\else
  \newcommand\listfigurename{List of Figures}
\fi
\ifdefined\listtablename
  \renewcommand*\listtablename{List of Tables}
\else
  \newcommand\listtablename{List of Tables}
\fi
\ifdefined\figurename
  \renewcommand*\figurename{Figure}
\else
  \newcommand\figurename{Figure}
\fi
\ifdefined\tablename
  \renewcommand*\tablename{Table}
\else
  \newcommand\tablename{Table}
\fi
}
\@ifpackageloaded{float}{}{\usepackage{float}}
\floatstyle{ruled}
\@ifundefined{c@chapter}{\newfloat{codelisting}{h}{lop}}{\newfloat{codelisting}{h}{lop}[chapter]}
\floatname{codelisting}{Listing}

\makeatother
\makeatletter
\@ifpackageloaded{caption}{}{\usepackage{caption}}
\@ifpackageloaded{subcaption}{}{\usepackage{subcaption}}
\makeatother

\ifLuaTeX
  \usepackage{selnolig}  
\fi
\usepackage[]{natbib}
\usepackage{bookmark}

\IfFileExists{xurl.sty}{\usepackage{xurl}}{} 
\hypersetup{
  pdftitle={Title},
  pdfauthor={Author 1; Author 2},
  pdfkeywords={3 to 6 keywords, that do not appear in the title},
  colorlinks=true,
  linkcolor={blue},
  filecolor={Maroon},
  citecolor={Blue},
  urlcolor={Blue},
  pdfcreator={LaTeX via pandoc}}

\usepackage{enumitem}
\usepackage{bm}
\usepackage{amsthm} 
\usepackage{etoolbox} 
\usepackage{mathtools} 
\usepackage{booktabs, makecell, multirow, caption}
\usepackage{orcidlink}
\usepackage{authblk}
\usepackage{setspace}

\newcommand{\T}{^{\top}}
\newcommand{\inv}{^{-1}}

\DeclareMathOperator{\Cov}{Cov}
\DeclareMathOperator{\Tr}{tr} 

\newcommand{\bome}{\bm{w}}
\newcommand{\bomeast}{{\bome^{\ast}}}
\newcommand{\hatbomeast}{\widehat{{\bome}^{\ast}}}

\newcommand{\bOmega}{\bm{\Omega}}
\newcommand{\bmu}{{\bm{\mu}}}
\newcommand{\hatbmu}{{\widehat{\bmu}}}

\newcommand{\bSig}{{\bm{\Sigma}}}

\newcommand{\hatbSig}{{\widehat{\bm{\Sigma}}}}

\newcommand{\bA}{{\bm{A}}}

\newcommand{\ba}{{\bm{a}}}

\newcommand{\bD}{\bm{D}}

\newcommand{\bI}{\bm{I}}

\newcommand{\NormDis}{\mathcal{N}}

\newcommand{\bQ}{{\bm{Q}}}
\newcommand{\bR}{{\bm{R}}}

\newcommand{\br}{\bm{r}}

\newcommand{\bs}{{\bm{s}}}

\newcommand{\bars}{{\bar{s}}}

\newcommand{\tildebs}{{\tilde{\bs}}}

\newcommand{\bX}{\bm{X}}

\newcommand{\by}{{\bm{y}}}

\newcommand{\bZ}{\bm{Z}}
\newcommand{\bz}{{\bm{z}}}

\DeclareMathOperator*{\argmax}{arg\,max}

\newtheorem{theorem}{Theorem}
\newtheorem{proposition}{Proposition}

\newtheorem{lemma}{Lemma}

\newtheorem{remark}{Remark}

\newtheorem{assumption}{Assumption}

\newcounter{assumptionInnerB}

\newcounter{assumptionInnerC}

\newcounter{assumptionInnerD}

\numberwithin{equation}{section}

\begin{document}

\title{Sustainable Investment: ESG Impacts on Large Portfolio}
\author[a,b]{Ruike Wu\orcidlink{0009-0000-3128-6214}}
\author[c]{Yonghe Lu}
\author[c]{Yanrong Yang\orcidlink{0000-0002-3629-5803}}

\affil[a]{Shanghai University of Finance and Economics}
\affil[b]{Key Laboratory of Mathematical Economics (Shanghai University of Finance and Economics), Ministry of Education}
\affil[c]{The Australian National University}

\date{}

\maketitle
\onehalfspacing
\begin{abstract}
This paper investigates the impact of environmental, social, and governance (ESG) constraint on a regularized mean-variance (MV) portfolio optimization problem in a large-dimensional setting, in which a positive definite regularization matrix is imposed on the sample covariance matrix. We first derive the asymptotic results for the out-of-sample (OOS) Sharpe ratio (SR) of the proposed portfolio, which help quantify the impact of imposing an ESG-level constraint as well as the effect of estimation error arising from the sample mean estimation of the assets’ ESG score. Furthermore, to study the influence of the choices of the regularization matrix, we develop an estimator for the OOS Sharpe ratio. The corresponding asymptotic properties of the Sharpe ratio estimator are established based on random matrix theory. Simulation results show that the proposed estimators perform close to the corresponding  oracle level. Moreover, we numerically investigate the impact of various forms of regularization matrices on the OOS SR, which  provides useful guidance for practical implementation.
Finally, based on OOS SR estimator, we propose an adaptive regularized portfolio which uses the best regularization matrix yielding the highest estimated SR (among a set of candidates) at each decision node.
Empirical evidence based on the S\&P 500 index demonstrates that the proposed adaptive ESG-constrained portfolio achieves a high OOS SR while satisfying the required ESG level, offering a practically effective approach for sustainable investment.
\end{abstract}


\newpage

\onehalfspacing

\section{Introduction}

The mean-variance (MV) optimization, first introduced by \cite{markowitz1952portfolio}, is a fundamental
cornerstone in modern portfolio allocation theory. For a universe of $p$ assets, the investors try to choose an allocation $\bome^*$ to maximize their utility as follows:
\begin{align}
   \bome^* =  \underset{\bome}{\arg\max}\;\bome\T{\bmu}-\frac{\gamma}{2}\bome\T{\bSig}\bome, 
   \label{eq classic problem}
\end{align}
where $\bmu=\left(\mu_{1},\mu_{2},...,\mu_{p}\right)\T$ is the expected return of assets return vector $\br =
\left(r_{1},r_{2},...,r_{p}\right)\T$, and $\bSig$ is the corresponding covariance matrix, and $\gamma$ is a tuning parameter measuring the degree of risk aversion. Clearly, the solution $\bome^*$ of this  optimization problem  is $\bSig^{-1}\bmu/\gamma$, which theoretically achieves the maximum Sharpe ratio.

In recent years, motivated by sustainable development goals and rising pro-social preferences, socially responsible investment (SRI) has become a rapidly expanding field. Environmental, social and governance (ESG) concerns—often regarded as ethical and non-financial dimensions of firm performance—have received substantial attention in both academia and industry. A growing number of investors actively allocate capital toward ESG-oriented products, and green or sustainability-themed funds have experienced substantial inflows in global markets. Meanwhile, regulatory initiatives are increasingly pushing  investors to incorporate ESG criteria into their portfolio decisions.
Numerous studies incorporate ESG information into portfolio decision and explore the trade-off between sustainability and financial returns \citep{barber2021impact, pedersen2021responsible, makridis2023balancing, berg2024quantifying, lo2024quantifying, lauria2025environmental}. In the mean-variance framework, a widely adopted approach imposes a minimum ESG score constraint, requiring the portfolio average ESG score to exceed a target threshold $\bar{s}$ \citep{pedersen2021responsible, cesarone2022does, de2023esg, lo2025performance}. As such, investors then seek allocations that maximize their mean-variance utility while ensuring ESG level; see optimization problem \eqref{eq optimization ESG level only}. This mechanism embeds ethical intentions directly into the mean-variance portfolio decision process and is elegant mathematically.

However, the primary challenge lies in ensuring trustworthy implementation. In practice, the population parameters $\bmu, \bSig$ as well as assets' ESG mean level are unknown and must be replaced by their empirical estimators. In modern markets, investors often manage hundreds or even thousands of tradable assets while only having access to limited historical observations. As a result, ESG-constrained MV optimization is commonly performed in large- or high-dimensional environments, in which the sample estimators of $\bmu$ and $\bSig$ suffer from severe estimation error and become statistically unreliable \citep{ao2019approaching, ding2021high, fan2022, bodnar2022optimal, wu2025making, ao2025robust, Meng30092025}. Moreover, when the number of assets is larger than the number of historical observations, the sample MV portfolio may even fail. Ethical intentions may be undermined if portfolio performance is overly distorted by estimation errors, resulting in a poorer balance between sustainable objectives and financial outcomes, and raising concerns about responsible  decision-making in finance. Thus, ensuring both financial performance and ethical validity of ESG-based allocation under vast data conditions (particularly large assets number) requires statistically principled and trustworthy methodologies.


In this paper, we investigate a regularized ESG-constrained mean–variance portfolios, aiming to achieve high Sharpe ratio while satisfying sustainable investment requirements, in large-dimensional settings. Specifically, based on the sample version of ESG-constrained mean-variance problem,  we propose adding a positive-definite regularization matrix to sample covariance matrix. 
Although this operation originates from statistical reasoning to address the ill conditioning of sample covariance estimation, imposing regularization on the sample covariance matrix can be equivalently reformulated as a quadratic constraint on the portfolio weights. And when the covariance matrix of the assets’ ESG scores is employed as the regularization matrix, this constraint effectively places an upper bound on the variance of the portfolio’s ESG scores. For the regularized portfolio, we first derive the asymptotic results for the out-of-sample (OOS) Sharpe ratio (SR) of the proposed regularized portfolio based on random matrix theory, which quantifies the impacts of imposing an ESG-level constraint as well as the estimation error, arising from estimating the  unknown ESG score mean level, in large dimensional situation. In the special case where the regularization matrix is set to a large multiple of the population covariance matrix, the estimation error in the sample ESG mean score  contributes to an improvement in the OOS SR. As a result, we can safely use the sample ESG mean as a proxy for the unknown ESG score in large dimensional situation. Furthermore, the OOS SR performance depends on the choice of the imposed regularization matrix, obtaining a theoretical optimum is challenging in the large-dimensional setting, where the number of assets may diverge with the sample size. Thus, in order to study the impacts of the choice of the regularization matrix on the OOS  portfolio SR,  we propose an estimator for OOS SR of our newly proposed portfolio. As such, investors can estimate the OOS SRs of the proposed portfolio at each desicion node under various choices of the regularization matrix, and then select the one that delivers the highest Sharpe ratio.  The asymptotic consistency of SR estimator is also established.



Our simulation results show that the proposed OOS SR estimator works well and is close to the unknown oracle level under  large-dimensional situations when simulation setup is calibrated from real data.  Furthermore, we investigate the impacts of several specifications of regularization  matrix on portfolio performance. The results show that setting the regularization matrix equal to the unknown true covariance matrix yields the best performance in terms of both SR and ESG score; however, this choice is infeasible in practice. Among feasible candidates, using the identity matrix, the diagonal matrix of the sample covariance matrix, and the nonlinear shrinkage estimator of the covariance matrix proposed by \cite{ledoit2017nonlinear} leads to high OOS  SR and the required ESG score, beating the naive sample version of ESG-constrained mean-variance portfolio. As such, these feasible choices can serve as useful guidance for practical implementation. Regularization matrices constructed using ESG score information (e.g. using covariance matrix of ESG scores as regularization matrix) fail to deliver satisfactory Sharpe ratio performance, indicating that the existing covariance structure of assets’ ESG scores contributes little to improving  performance of sample ESG-constrained mean-variance portfolio  when an additional ESG variance constraint is further imposed.

Finally, an empirical analysis based on the components of the S\&P 500 index is conducted.  Due to the complexity of real data, we apply an adaptive version of the newly proposed portfolio: at each decision node, we first estimate the OOS SR for various candidate regularization matrices (in our study, the candidates include the identity matrix, the diagonal matrix of the sample covariance matrix, and the nonlinear shrinkage covariance matrix estimator), and then select the regularization matrix that exhibits the highest estimated Sharpe ratio. 
Compared with other feasible ESG-constrained strategies, the proposed adaptive regularized portfolio achieves first-tier OOS SR performance in most cases, while maintaining the ESG levels that satisfactorily meet the prescribed constraint and exhibit low variability.

This paper makes several contributions to the literature on sustainable investment and constrained large portfolio allocation:
\begin{itemize}
\item[(1)] We investigate a  regularized ESG-constrained mean–variance  optimization problem in large dimensional situation. We derive the limit of OOS SR of the proposed portfolio, which quantifies the impacts of imposing an ESG-level constraint as well as the estimation error from ESG mean level estimator in large dimensional situation. 

\item[(2)] 
We develop an estimator for the OOS SR of the regularized portfolio, which is useful for investors to select  the best forms of regularization matrix in terms of maximizing OOS SR. 
The asymptotic properties  are established based on random matrix theory. 
\item[(3)] 
We study the impact of various forms of regularization matrices on the performance of ESG-constrained mean-variance portfolios through simulation experiments, and conclude that,  using the identity matrix, the diagonal matrix of the sample return covariance matrix, and the nonlinear shrinkage covariance estimator leads to high OOS Sharpe ratios while satisfying the required ESG score, thereby providing useful guidance for empirical applications. 
\item[(4)]
We propose an adaptive regularized ESG-constrained mean-variance portfolio that uses the regularization matrix yielding the highest estimated OOS Sharpe ratio among a set of candidates at each decision node.  The empirical analyses based on S\&P 500 constituent data show that, the proposed adaptive portfolio achieves a high OOS SR while obtaining stable ESG scores satisfying the target level well. This provides a  practically effective approach for balancing financial performance with sustainability goals.
\end{itemize}

This paper is related to the studies that integrate ESG information into portfolio selection. 
\cite{utz2015tri} incorporate sustainability as a third criterion, in addition to the mean-variance trade-off, within a tri-criterion portfolio selection framework.
\cite{gasser2017markowitz},  \cite{pedersen2021responsible}, and \cite{schmidt2022optimal} incorporate ESG measure into portfolio optimizing utility function, and provide ESG-Sharpe ratio efficient frontier.  
\cite{makridis2023balancing} introduce ESG scores into portfolio allocation based on linear shrinkage covariance estimator \citep{ledoit2004well}, they study the situations where the target matrix is defined using the reciprocal
of the ESG ratings. \cite{de2023esg} examine the impact of including ESG criteria in the allocation of equity portfolios under mean-variance framework empirically. 
\cite{lo2025performance} propose a performance attribution framework for mean-variance optimal portfolio constraints, and study the financial impact of the ESG level constraint empirically.
\cite{lauria2025environmental} incorporate ESG scores into a dynamic asset pricing framework by constructing an ESG-valued log return, defined as a linear combination of the raw return and the scaled ESG scores. All these studies are confined to cases where the sample size substantially exceeds the number of dimensions, and do not examine (theoretically) the performance of constrained portfolios in high-dimensional settings.

This paper also connects to the regularized portfolio allocation and high dimensional portfolio allocation. \cite{demiguel2009generalized} developed a general framework that solves the global minimum variance problem (GMVP) by incorporating the A-norm constraint for
the portfolio weight vector. \cite{fan2013large} proposed a latent factor–based covariance estimation method, and the underlying idea of employing factor models to alleviate the curse of dimensionality has been extensively adopted in portfolio allocation research.   \cite{ding2021high} developed a unified GMVP under statistical factor models in high-dimensional situations, \cite{fan2022} proposed a time-varying minimum variance portfolio by letting time-varying factor loading, and \cite{wu2025making} extend the famous distributionally-robust mean-variance portfolio of \cite{blanchet2022distributionally} into high dimensional situation based on factor structure.
\cite{ao2019approaching} studied the mean–variance optimization with an $\ell_1$-norm regularization by reformulating it as a regression problem in high dimensional situation, and this approach has been further generalized to accommodate high-dimensional settings where dimension is larger than sample size \citep{li2022synthetic, chen2025cross}, as well as heteroskedastic and heavy-tailed data \citep{ao2025robust}. \cite{Meng30092025} proposed to correct the
sample covariance matrix by adding a regularization within the framework of mean-variance problem,  in high dimensional setup, and provided a consistent estimator of
OOS SR of the regularized portfolio. All the aforementioned studies focus solely on the mean-variance problem and do not incorporate additional information such as ESG scores for non-financial objectives.

This paper is organized as follows: 
Section~\ref{Sec:Problem} introduces the regularized ESG-constrained mean-variance portfolio. Section~\ref{sec: asym} develops the asymptotic results for the OOS SR of the regularized portfolio, proposes  OOS SR estimator and establishes its asymptotic properties. Section~\ref{sec: simu} presents simulation studies. Section~\ref{sec: empirical} proposes adaptive regularized ESG-constrained mean-variance portfolio, and conducts  empirical analysis based on the components of the S\&P 500 index. Section~\ref{sec: conclude} concludes the paper. 

In this paper, $\| \bA\|$, $\|\bA\|_1$, $\|\bA\|_F$, $\|\bA\|_{\infty}$ and $\|\bA\|_{\textup{tr}}$ denote the spectral norm, $L_1$ norm, Frobenius norm, max norm and trace norm of a matrix $\bA$, defined respectively by $\|\bA\| = \lambda_{\max}^{1/2}(\bA^\top \bA)$, $\|\bA\|_1 = \max_{j}\sum_{i} |\bA_{ij}|$, $\|\bA\|_F = \text{tr}^{1/2}(\bA^\top \bA)$, $\|\bA\|_{\infty} = \max_i \sum_{j} |\bA_{ij}|$ and $\|\bA\|_{\textup{tr}}=\Tr[(\bA\T\bA)^{1/2}]$, 
$\lambda_{\max}(\bA)$ ($\lambda_{\min}(\bA)$) denotes the maximum (minimum) eigenvalues of a matrix $\bA$ and $\text{tr}(\bA)$ is the trace of a matrix $\bA$. If $\bA$ is a column vector, both  $\|\bA\|$ and $\|\bA\|_F$ equal the Euclidean norm $\|\bA\|_2$, and $\|\bA\|_{\infty} = \max_{i}|\bA_{i}|$. $a_T\asymp b_T$ means that for some constants~$c, C > 0$, $c<|a_T|/|b_T|<C$ for all $T$.

\section{Regularized ESG-Constrained Problem}\label{Sec:Problem}

Consider the following portfolio optimization problem, which regulates the portfolio’s ESG score to  $\bars$:\footnote{Without loss of generality, we consider the equality constraint on the ESG level, rather than the inequality one.} \begin{eqnarray}
\underset{\bome}{\arg\max}\;\bome\T{\bmu}-\frac{\gamma}{2}\bome\T{\bSig}\bome \qquad \text{s.t.}\quad \bome\T \bs/\bome^\top \bm{1}  = \bars,
\label{eq optimization ESG level only}
\end{eqnarray}
where $\bome\T \bs/\bome^\top \bm{1}  = \bars$ is
the sustainability requirement (hereafter, ESG level constraint) on standardized portfolio ESG score, and $\bs = \textbf{E}(\ba_t)$ is the expected ESG score of considered assets, and $\ba_t$ is the ESG score vector.\footnote{The sustainable variable, the ESG score, can be replaced by other variables (e.g., carbon emissions, sustainability) for impact investing purposes.} 
Define $\tildebs:=\bs-\bm{1}\bars$, then the ESG level constraint can be rewritten as $\bome^\top \tildebs = 0$.   The constrained optimization problem \eqref{eq optimization ESG level only} has the following  optimal solution in closed-form:
\begin{equation}
\label{Eq: ESG level only optimal_solution}
\bomeast=\frac{1}{\gamma}{\bSig}\inv\left({\bmu}+{\xi}{\tildebs}\right),
\end{equation}
where $\xi  = -{\bmu\T\bSig\inv\tildebs}/{\tildebs\T\bSig\inv\tildebs}.$  A straightforward calculation shows that the OOS Sharpe ratio of the oracle portfolio $\bome^\ast$ is given by 
\begin{align}
\theta_{max} = \sqrt{\bmu^\top \bSig^{-1}\bmu - \frac{(\bmu\T\bSig\inv\tildebs)^2}{\tildebs\T\bSig\inv\tildebs}}.
\label{eq SR max}
\end{align}
From the oracle-population perspective, imposing the ESG constraint strictly reduces the Sharpe ratio of the optimal allocation unless $\bmu^\top \bSig^{-1}\tildebs = 0$, given that $\bSig^{-1}$ is positive definite.
The oracle Sharpe ratio $\theta_{\max}$ serves as a theoretical upper bound for estimated ESG-constrained portfolios.

 In practical application, due to the lack of information of $\bmu$, $\bSig$, and $\tildebs$,  the practical optimization using sample estimations is considered as follows:  
\begin{align}
\label{eq: empirical version}
\underset{\bome}{\arg\max}\;\bome\T\hatbmu-\frac{\gamma}{2}\bome\T\hatbSig\bome \qquad \text{s.t.}\quad \bome\T\widehat{\tildebs}=0,
\end{align}
where $\hatbmu, \hatbSig$ and $\widehat{\tildebs}$ are sample estimations calculated based on collected data. 
By using the first-order arguments, the  optimization \eqref{eq: empirical version} is equivalent to the following non-constrained problem:
\begin{align}
\label{eq: empirical version 2}
\underset{\bome}{\arg\max}\;\bome\T\left(\hatbmu + \widehat{\xi} \widehat{\tildebs} \right)-\frac{\gamma}{2}\bome\T\hatbSig\bome,
\end{align}
where $\widehat{\xi}  = -{\widehat{\tildebs}\T\hatbSig\inv\hatbmu}/{\widehat{\tildebs}\T\hatbSig\inv\widehat{\tildebs}}.$
From \eqref{eq: empirical version 2}, the ESG level constraint essentially penalizes the sample mean estimation, which is known to be unreliable in large dimensional settings \citep{demiguel2009generalized,ao2019approaching,bodnar2019optimal}. Therefore, $\widehat{\tildebs}$ serves as a penalization term on the sample mean estimation. 
Furthermore, it is also well known that the empirical covariance estimator performs poorly in large dimensional situations, particularly in the filed of portfolio allocation \citep{michaud1989markowitz, jagannathan2003risk, kan2007optimal, bodnar2022optimal, lassance2024risk}. As such,
in order to mitigate the bad condition number of the sample covariance estimation,  particularly when dimension is larger than sample size, similar to formulation of the sample mean part, we consider adding an regularization matrix term $\bQ$ on the sample covariance \citep{ledoit2003improved, demiguel2009generalized, makridis2023balancing, Meng30092025}, and thus consider the following problem:
\begin{align}
\label{eq: empirical version 3}
\underset{\bome}{\arg\max}\;\bome\T\hatbmu-\frac{\gamma}{2}\bome\T\left(\hatbSig+\eta \bQ\right)\bome, \quad \bome^\top \widehat{\tildebs} = 0,
\end{align}
where $\bQ$ is a positive definite deterministic square matrix that determines the direction of regularization, and the coefficient $\eta$ controls the degree of penalization. 
Similarly,  the optimization problem \eqref{eq: empirical version 3} has a closed-form optimal solution as follows:
\begin{equation}
\label{Eq:optimal_solutionnew logic}
\hatbomeast=\frac{1}{\gamma}\left(\hatbSig+\eta \bQ\right)\inv\left(\hatbmu+\widehat{\zeta}\widehat{\tildebs}\right),
\end{equation}
where $\widehat{\zeta}(\eta)  = -{\widehat{\tildebs}\T(\hatbSig+\eta \bQ)\inv \hatbmu}/{\widehat{\tildebs}\T(\hatbSig+\eta \bQ)\inv\widehat{\tildebs}}$. It is worth noting that although adding a well-behaved regularization term to address poor conditioning is common in high-dimensional statistical inference, applying such a regularization to portfolio optimization also carries a meaningful economic interpretation in the context of ESG-based portfolio allocation.

\begin{remark} 
\label{remark economic}
By simple calculation shown in our online Appendix~A, the optimal solution of  problem \eqref{eq: empirical version 3} is equivalent to the solution to following optimization 
\begin{align}
\underset{\bome}{\max}\;\bome\T\hatbmu-\frac{\gamma}{2}\bome\T\hatbSig\bome \qquad \text{s.t.}\quad \bome\T\widehat{\tildebs}=0,\;\;\bome\T \bQ \bome =  \delta(\eta),
\label{eq ESG problem}
\end{align}
where $\delta(\eta) = \left(\hatbmu+\widehat{\zeta}(\eta)\widehat{\tildebs}\right)^\top \left(\hatbSig+\eta \bQ\right)\inv \bQ\left(\hatbSig+\eta \bQ\right)\inv\left(\hatbmu+\widehat{\zeta}(\eta)\widehat{\tildebs}\right)/\gamma^2$.  Adding a regularization matrix to the sample covariance matrix is equivalent to imposing an additional constraint that the A-norm of the portfolio weight  equals a threshold $\delta(\eta)$ \citep{demiguel2009generalized}. In particular,  if $\bQ = \bOmega$ where $\bOmega$ is the covariance matrix of assets' ESG scores,  the quadratic constraint $\bome^\top \bOmega \bome = \delta(\eta)$ requires that the variance of portfolio ESG scores is restricted to the target level $\delta$.
Therefore, the optimization problem \eqref{eq: empirical version 3} maximizes the investor's utility subject to two sustainability constraints on the portfolio’s ESG average level and variance. Clearly, an appropriate specification of 
$\bQ$ can provide valuable guidance for the design and quantification of ESG scores, enabling investors to achieve superior performance while maintaining sustainability requirements. Furthermore, when $\bQ = \bI_p$,  there is a one-to-one correspondence between the A-norm-constrained portfolios and unconstrained portfolios computed based on the linear shrinkage covariance matrix \citep{ledoit2004well}. 
\end{remark}

\begin{remark}
The studies of \cite{hu2025sustainability} and \cite{lauria2025environmental} integrate the ESG information into decision-making by using the  ESG-valued return $\br_t^{esg}$, which is defined as $\br_t^{esg} = (1-\iota) \br_t +  \iota \tilde{\ba}_t$ where $\tilde{\ba}_t$ is the scaled ESG score and $\iota \in (0,1)$ is the intensity of ESG preference. Note that $\textbf{E}(\br_t^{esg}) = (1-\iota)\left(\bmu +\frac{\iota}{1-\iota}\textbf{E}(\tilde{\ba}_t)\right)$ and $\Cov(\br_t^{esg}) = (1-\iota)^2\left(\bSig+\frac{\iota^2}{(1-\iota)^2}\tilde{\bOmega}\right)$ if $\tilde{\ba}_t$ and $\br_t$ are uncorrelated, and $\tilde{\bOmega}$ is the covariance matrix of $\tilde{\ba}_t$. As a result, when the ESG-valued return is applied to mean-variance problem \eqref{eq classic problem} with $\tilde{\ba}_t =  \ba_t -  \bm{1}\bars$ and risk aversion parameter $\gamma = \gamma_1/(1-\iota)$, it is equivalent to solving the population version of the regularized problem \eqref{eq: empirical version 3}  with $\gamma =\gamma_1$, $\bQ = \bOmega$, and $\eta = \zeta(\eta)^2 = (\iota/(1-\iota))^2$.
\end{remark}

\begin{remark}
   When $\bQ$ is a diagonal matrix with positive diagonal entries, the term $\bome^\top \left(\frac{\eta\gamma}{2}\bQ\right)\bome$ can be interpreted as explicitly accounting for transaction costs; see, e.g., \cite{hautsch2019large} and \cite{fan2024cost}. As a result, with an appropriately selected penalization matrix $\bQ$, our model promotes more stable allocations.
\end{remark}

Based on optimal solution \eqref{Eq:optimal_solutionnew logic}, the OOS SR of estimated regularized ESG-constrained mean-variance  can be further defined as
\begin{equation}
\theta^\ast(\eta,\bQ):=\frac{\left(\hatbmu+\widehat{\zeta}\widehat{\tildebs}\right)\T(\hatbSig+\eta \bQ)^{-1}\bmu}{\sqrt{\left(\hatbmu+\widehat{\zeta}\widehat{\tildebs}\right)\T(\hatbSig+\eta \bQ)^{-1}\bSig(\hatbSig+\eta \bQ)^{-1}\left(\hatbmu+\widehat{\zeta}\widehat{\tildebs}\right)}}.
\label{eq SR}
\end{equation}
From \eqref{eq SR}, it is evident that $\theta^\ast(\eta,\bQ)$ depends on the choice of $\bQ$ and $\eta$, and in what follows we drop $(\eta,\bQ)$ whenever this dependence is clear from the context. In the next section, we study the asymptotic properties of the OOS SR of the estimated ESG-constrained portfolio, which provides insight into how the ESG constraint affects portfolio performance in large-dimensional settings.

\section{Asymptotic Analysis}\label{sec: asym}

In this section, we conduct an asymptotic analysis of the out-of-sample (OOS) Sharpe ratio (SR) for ESG-constrained  portfolios. Before deriving the asymptotic behavior of ${\theta}^\ast(\eta,\bQ)$, we first introduce some required assumptions.



\begin{assumption}
\label{assum:pT_ratio}
We consider an asymptotic setup where $T, p\rightarrow\infty$, and $ p/T\rightarrow c\in(0,1)\cup (1,\infty)$.
\end{assumption}

Assumption \ref{assum:pT_ratio} is quite standard in the literature of random matrix theory. We exclude the knife-edge case $c = 1$ because the sample eigenvalue density diverges near zero in this regime, leading to additional technical complications \citep{ledoit2011eigenvectors,ledoit2017nonlinear,lu2024double}.

\begin{assumption}
\label{assum:data_stock_return RE}
The excess returns vector at time $t$ is generated as $\br_t=\bmu+\bSig^{\frac{1}{2}}\bz_t$, where $\bz_t\stackrel{\text{i.i.d.}}{\sim}\NormDis\left(\bm{0},\bI_p\right)$.
\end{assumption}

Gaussianity assumption for return data is common and standard in random-matrix-based portfolio theory; see, for example, \cite{el10}, \cite{bodnar2022recent}, \cite{li2022spectrally} and \cite{bodnar2024two}. 
Stacking the $T$ observations yields the asset return matrix $\bR=\bm{1}_T \bmu^\top+\bX$, where $\bm{1}_T$ is a $T$-dimensional column vector of ones and $\bX = \bZ\bSig^{\frac{1}{2}}$, with $\bZ=[\,\bz_1^\top,\ldots,\bz_T^\top\,]^\top \in \mathbb{R}^{T\times p}$. Under this representation, the sample covariance matrix is $\hatbSig := \bX^\top \bX/T$.


\begin{assumption}
\label{assum:Sigma} The matrix $\bSig$ is well scaled as $\lVert\bSig/p\rVert_{\textup{tr}}\leq C$ for some constant $C>0$. Denote the eigenvalues of $\bSig$ as $\lambda_1\geq\cdots\geq\lambda_p$, one of the following conditions hold:\\
1. When $p<T$, we allow arbitrary number of diverging eigenvalues.\\
2. When $p>T$, we allow at most a fixed number $K$ of diverging eigenvalues, and $\lambda_1\leq C\lambda_K^2$ for some constant $C>0$.
\end{assumption}

Both cases accommodate factor-type covariance structures with spiked eigenvalues that may diverge as $p$ grows. The distinction between Cases~1 and~2 is mainly technical and depends on the dimensionality regime. 

\begin{assumption}
\label{assum:Q} 
The positive definite regularization matrix $\bQ\in \mathcal{Q}$, independent of $\bX$, satisfies the following: there exist constants $c_1\geq$, $c_2>0$, with $c_1^2+c_2^2>0$, and $\bQ_1$ and $\bQ_2$ such that $\bQ=c_1\bQ_1+c_2\bQ_2$. The matrices $\bQ_1$ and $\bQ_2$ satisfies $c'\bI\preceq\bQ_1\preceq C'\bI$ and $c'\bI\preceq\bSig^{-1/2}\bQ_2\bSig^{-1/2}\preceq C'\bI$ for some constant $c', C'>0$.
\end{assumption}

Assumption~\ref{assum:Q} rules out ill-conditioned choices of the regularization matrix $\bQ$.

\begin{assumption}
\label{assum:data_esg}
The ESG score vector $\ba_t$ corresponding to the $p$ assets at time $t$ is generated as $\ba_t = \bs + \bOmega^{\frac{1}{2}}\by_t$, where $\by_t\stackrel{\text{i.i.d.}}{\sim}\NormDis\left(\bm{0},\bI_p\right)$. The $\bOmega$ has uniformly bounded spectral norm; that is, $\lambda_{\max}(\bOmega)\leq C$ for some constant $C>0$. In addition, $\bs\T\bSig\inv\bs, \bs\T\bSig\inv\bm{1}, \bm{1}^\top\bSig\inv\bm{1} $ are bounded, and $\ba_t$ is assumed to be independent of the return  $\br_t$.
\end{assumption}

Assumption \ref{assum:data_esg} postulates that the asset-level ESG score vector $\ba_t$ follows a multivariate normal distribution, which serves as a technical requirement for the random matrix theory framework. 
When the covariance matrix $\bOmega$ is close to the zero matrix, $\ba_t$ can be treated as approximately time-invariant. The condition that $\bm{1}^\top \bSig^{-1} \bm{1}$ is bounded holds when the minimum risk is non-diversifiable; see Example 2.2 of \cite{ding2021high}. The boundedness of $\bs^\top \bSig^{-1} \bs$ and $\bs^\top \bSig^{-1} \bm{1}$ is mild, given that ESG scores are measured on arbitrary, provider-specific scales and can be standardized prior to evaluation. Based on Assumption \ref{assum:data_esg}, the sample mean of ESG score is $\widehat{\bs} := \frac{1}{T}\sum_{t=1}^T \ba_t$,
and the centered ESG exposure is $\widehat{\tildebs} = \widehat{\bs} - \bm{1}\,\bars$. For technical conciseness, the ESG scores and returns are assumed to be independent. We also conduct statistical analysis on the real datasets and find that, the ESG scores and returns indeed have weak correlations.




The following quantities will be used repeatedly in subsequent analysis. For notational convenience, we first define 
$\mathbb{G}
:=\left(\frac{\bSig}{1+s_{0}}+\eta\bQ\right)^{-1}$and $\mathbb{H}
:=\left(\frac{\bSig}{1+s_{0}}+\eta\bQ\right)^{-1}\bSig
\left(\frac{\bSig}{1+s_{0}}+\eta\bQ\right)^{-1},
$
where $s_{0}$ is defined as the unique solution to
$
s_{0}=\frac{c}{p}\Tr\left[\bSig\left(\frac{\bSig}{1+s_{0}}+\eta\bQ\right)\inv\right],
$
and $s_{1,\bSig}$ is given by
$
s_{1,\bSig}=\left(\frac{s_{1,\bSig}}{(1+s_{0})^2}-1\right)\frac{c}{p}\Tr\left[\bSig\left(\frac{\bSig}{1+s_{0}}+\eta\bQ\right)\inv\bSig\left(\frac{\bSig}{1+s_{0}}+\eta\bQ\right)\inv\right].
$
Under these conditions, the following lemma characterizes the limit of OOS Sharpe ratio $\theta^\ast$ of proposed ESG-constrained portfolio.

\begin{lemma}
\label{lemma OOS SR asymtotic}
Suppose Assumptions \ref{assum:pT_ratio}-\ref{assum:data_esg} hold true, we have 
\begin{align}
{\theta}^* -   
\frac{ \bmu\T\mathbb{G}\bmu - \frac{\left(\bmu\T\mathbb{G}\tildebs\right)^2}{\tildebs\T\mathbb{G}\tildebs + T^{-1}\Tr\left[\mathbb{G}\bOmega\right]}}{{\sqrt{1-\frac{s_{1,\bSig}}{(1+s_0)^2}}}\sqrt{\bmu\T\mathbb{H}\bmu+T^{-1}\Tr\left[\mathbb{H}\bSig\right]  +
\frac{\left(\bmu\T\mathbb{G}\tildebs\right)^2 \left(\tildebs\T\mathbb{H}\tildebs + T^{-1}\Tr\left[\mathbb{H}\bOmega\right]\right)}{\left(\tildebs\T\mathbb{G}\tildebs + T^{-1}\Tr\left[\mathbb{G}\bOmega\right]\right)^2}  -  2 \frac{\bmu\T\mathbb{G}\tildebs \bmu\T\mathbb{H}\tildebs}{\tildebs\T\mathbb{G}\tildebs + T^{-1}\Tr\left[\mathbb{G}\bOmega\right]}}}\overset{a.s}{\longrightarrow}0.  \label{eq theta ast converge} 
\end{align}
\end{lemma}

Lemma \ref{lemma OOS SR asymtotic} shows the limit of OOS SR of the newly proposed  regularized ESG-constrained mean-variance portfolio \eqref{Eq:optimal_solutionnew logic}. The limit of $\theta^\ast$ is complicated. Thus, in order to better analyze the asymptotic behavior of  OOS SR,  we further  introduce the following mixed-information regularized  optimization problem which relies on population ESG information $\tildebs$, while using the sample mean and   sample covariance matrix of assets return:
 \begin{align}
\label{eq: Mix problem constraint}
\underset{\bome}{\arg\max}\;\bome\T\hatbmu-\frac{\gamma}{2}\bome\T\left(\hatbSig+\eta \bQ\right)\bome, \quad \bome^\top {\tildebs} = 0.
\end{align}
The solution to \eqref{eq: Mix problem constraint} is given by $\check{\bome} =\left(\hatbSig+\eta \bQ\right)\inv\left(\hatbmu+\tilde{\zeta}{\tildebs}\right)/\gamma$, where $\check{\zeta}$ is defined as  $\check{\zeta}  = -{{\tildebs}\T(\hatbSig+\eta \bQ)\inv \hatbmu}/{{\tildebs}\T(\hatbSig+\eta \bQ)\inv{\tildebs}}$, and thus the corresponding OOS SR of $\check{\bome}$ is:
\begin{equation}
\check{\theta}^\ast(\eta,\bQ):=\frac{\left(\hatbmu+\check{\zeta}{\tildebs}\right)\T(\hatbSig+\eta \bQ)^{-1}\bmu}{\sqrt{\left(\hatbmu+\check{\zeta}{\tildebs}\right)\T(\hatbSig+\eta \bQ)^{-1}\bSig(\hatbSig+\eta \bQ)^{-1}\left(\hatbmu+\check{\zeta}{\tildebs}\right)}}.
\label{eq SR ESG know}
\end{equation}

\begin{lemma}
\label{lemma SR asymtotic mix 2}
    Suppose Assumptions \ref{assum:pT_ratio}-\ref{assum:data_esg} hold true, we have 
\begin{align}
\check{\theta}^\ast -  
\frac{ \bmu\T\mathbb{G}\bmu -\frac{\left(\bmu\T\mathbb{G}\tildebs\right)^2}{\tildebs\T\mathbb{G}\tildebs}}{{\sqrt{1-\frac{s_{1,\bSig}}{(1+s_0)^2}}}\sqrt{\bmu\T\mathbb{H}\bmu+T^{-1}\Tr\left[\mathbb{H}\bSig\right]  +
\frac{\left(\bmu\T\mathbb{G}\tildebs\right)^2 \tildebs\T\mathbb{H}\tildebs}{\left(\tildebs\T\mathbb{G}\tildebs\right)^2}  -  2 \frac{\bmu\T\mathbb{G}\tildebs \bmu\T\mathbb{H}\tildebs}{\tildebs\T\mathbb{G}\tildebs}}}\overset{a.s}{\longrightarrow}0. \label{eq check theta ast converge} 
\end{align}
\end{lemma}

Lemma \ref{lemma SR asymtotic mix 2} presents the asymptotic result of $\check{\theta}^{\ast}$.   Compare the limits in  Lemmas \ref{lemma OOS SR asymtotic} and \ref{lemma SR asymtotic mix 2},
the impacts of estimation error from estimating ESG scores mean level in high dimensional situation can be quantified. Specifically, it is evident that,  the estimation error of sample mean of ESG scores introduces an additional term $\Tr\left[\mathbb{G}\bOmega \right]/T$ (or $\Tr\left[\mathbb{H}\bOmega \right]/T$) to the term $\tildebs^\top \mathbb{G}\tildebs$ (or $\tildebs^\top \mathbb{H}\tildebs$), and thus affect the financial performance of estimated portfolio. Clearly, when assets’ ESG scores are  stable enough so that $\bOmega$ is close to zero matrix $\bm{0}_{p\times p}$, the limit in \eqref{eq theta ast converge} degenerates to that in \eqref{eq check theta ast converge}.

\begin{remark}
Let $\tilde{\bome}$ denote the solution of $\underset{\bome}{\max}\;\bome\T\bmu-\frac{\gamma}{2}\bome\T\left(\hatbSig+\eta \bQ\right)\bome$ subject to $\bome^\top {\tildebs} = 0$. For this optimization problem, {only sample covariance matrix is applied}, and population mean of return and ESG scores are considered as known.
 By using the similar arguments of Lemma \ref{lemma SR asymtotic mix 2}, the OOS SR of $\tilde{\bome}$ (denoted as $\tilde{\theta}^{\ast}$) has a limit of $$\lim_{T\rightarrow\infty}\frac{ \bmu\T\mathbb{G}\bmu -\frac{\left(\bmu\T\mathbb{G}\tildebs\right)^2}{\tildebs\T\mathbb{G}\tildebs}}{{\sqrt{1-\frac{s_{1,\bSig}}{(1+s_0)^2}}}\sqrt{\bmu\T\mathbb{H}\bmu  +
\frac{\left(\bmu\T\mathbb{G}\tildebs\right)^2 \tildebs\T\mathbb{H}\tildebs}{\left(\tildebs\T\mathbb{G}\tildebs\right)^2}  -  2 \frac{\bmu\T\mathbb{G}\tildebs \bmu\T\mathbb{H}\tildebs}{\tildebs\T\mathbb{G}\tildebs}}}. $$  A similar result for unconstrained mean-variance portfolio can be referred to  \cite{Meng30092025}. As such, Lemma \ref{lemma SR asymtotic mix 2} also  quantifies the impact of the estimation error of the sample mean in high-dimensional situations, i.e., introducing an additional term $\Tr\left[\mathbb{H}\bSig\right]/T$. It is worth noting that, under the conditions of Lemma \ref{lemma SR asymtotic mix 2}, if $\bQ = C\bSig$ with a sufficiently large constant $C$, the  $\tilde{\theta}^\ast$ converges to the theoretical maximum Sharpe ratio $\theta_{\max}$ defined in \eqref{eq SR max}. As such, with an appropriate choice of $\bQ$, it is possible to approximate the theoretical maximum Sharpe ratio in high-dimensional settings when only sample covariance matrix is applied.  However, when sample mean is further applied (i.e. problem \eqref{eq: Mix problem constraint}), designing $\bQ$ to be proportional to $\bSig$ cannot attain the theoretical maximum due to the presence of $\Tr\left[\mathbb{H}\bSig\right]/T$.
\label{Remark Meng 2025}
\end{remark}



\begin{remark}
\label{remark unconstrained port}
 Let $\dot{\bome}$ denote the solution of $\underset{\bome}{\max}\;\bome\T\hatbmu-\frac{\gamma}{2}\bome\T\left(\hatbSig+\eta \bQ\right)\bome$ without ESG level constraint,  the limit of OOS SR of $\dot{\bome}$ is $\lim_{T\rightarrow\infty}\frac{ \bmu\T\mathbb{G}\bmu}{{\sqrt{1-\frac{s_{1,\bSig}}{(1+s_0)^2}}}\sqrt{\bmu\T\mathbb{H}\bmu+\frac{1}{T}\Tr\left[\mathbb{H}\bSig\right] }} $. Compare to \eqref{eq check theta ast converge}, it is clear that imposing the ESG level constraint reduces the value of the numerator. The impact on the denominator, however, depends on the specific choice of the regularization matrix. In the special case where $\bQ = C\bSig$ with sufficiently large $C$, the limiting OOS SR of $\dot{\bome}$ can be rewritten as $\frac{\bmu^\top \bSig^{-1}\bmu}{\sqrt{\bmu^\top \bSig^{-1}\bmu + c}}$, and that of $\check{\bome}$ (optimal portfolio with ESG level constraint) is
$\frac{\bmu^\top \bSig^{-1}\bmu - \frac{(\bmu\T\bSig\inv\tildebs)^2}{\tildebs\T\bSig\inv\tildebs}}{\sqrt{\bmu^\top \bSig^{-1}\bmu + c - \frac{(\bmu\T\bSig\inv\tildebs)^2}{\tildebs\T\bSig\inv\tildebs}}}$, see Proposition~\ref{prop 2} below. Since the function $f(x) = x/\sqrt{x+c}$ is increasing in $x$, imposing the ESG constraint leads to a decrease in the OOS SR.
\end{remark}

A general comparison without specifying the form of $\bQ$ is analytically intractable. 
As such, we now focus on a special situation where $\bQ = C\bSig$ with sufficiently large $C$, which is the optimal choice for the regularized  mean-variance  problem using sample covariance matrix but population mean return and  mean ESG score, as discussed in Remark \ref{Remark Meng 2025}.
\begin{proposition}
\label{prop 2}
Under the conditions of Lemma \ref{lemma OOS SR asymtotic}, when $\bQ = C\bSig$, where $C$ is allowed to depend on $T$ and satisfies $C\to\infty$ as $T\to\infty$,
we have
\begin{align*}
 &\check{\theta}^*-\frac{\bmu^\top \bSig^{-1}\bmu - \frac{(\bmu\T\bSig\inv\tildebs)^2}{\tildebs\T\bSig\inv\tildebs}}{\sqrt{\bmu^\top \bSig^{-1}\bmu + c - \frac{(\bmu\T\bSig\inv\tildebs)^2}{\tildebs\T\bSig\inv\tildebs}}}\overset{a.s}{\longrightarrow}0,
 \\ &{\theta}^* -\frac{\bmu^\top \bSig^{-1}\bmu - \frac{(\bmu\T\bSig\inv\tildebs)^2}{\tildebs\T\bSig\inv\tildebs+T^{-1}\Tr\left[\bSig\inv\bOmega\right]}}{\sqrt{\bmu^\top \bSig^{-1}\bmu + c - \frac{(\bmu\T\bSig\inv\tildebs)^2}{\tildebs\T\bSig\inv\tildebs+T^{-1}\Tr\left[\bSig\inv\bOmega\right]}}}\overset{a.s}{\longrightarrow}0.
\end{align*}
\end{proposition}
First, for both $\check{\bome}$ and $\hatbomeast$, their OOS SR cannot attain the theoretical maximum $\theta_{\max}$ due to the presence of the term $c=\lim_{T\to\infty} p/T$ in the denominator. Nevertheless, our simulation results indicate that setting $\bQ=\bSig$ remains a good choice, as it yields a higher OOS SR than the other considered alternatives of $\bQ$ under finite sample. Second, interestingly, the estimation error in the ESG score mean does not adversely affect the financial results and, in fact, leads to higher Sharpe ratio performance. Specifically, let $A := \bmu^\top \bSig^{-1}\bmu$, $B := \bmu^\top \bSig^{-1}\tildebs$.
For function $f(x) = (A - B^2/x)/\sqrt{A+c-B^2/x}$, by simple algebra, it is derivative with respect to $x$ is ${B^2}(A+2c -{B^2}/{x})/\left(2x^2\right(A+c-\frac{B^2}{x})^{3/2})$, which is strictly positive at $x = \tildebs^\top \bSig^{-1}\tildebs$, given  that $(\bmu^\top \bSig^{-1}\bmu)(\tildebs^\top \bSig^{-1}\tildebs) \geq (\bmu^\top \bSig^{-1}\tildebs)^2$ from Cauchy-Schwarz inequality. {As such, a positive value of $\Tr[\bSig^{-1}\bOmega]$ implies that the estimation error from sample ESG scores has a favorable impact on the OOS SR, whereas a negative value indicates a negative effect on portfolio performance.}
Since $\bSig$ and $\bOmega$ are covariance matrices, which are symmetric and assumed to be positive definite, it is evident that the estimation error from estimating ESG score mean lead to an improvement in SR. As a result, we can safely use the sample ESG mean estimation as a proxy for the unknown the ESG mean level in large dimensional situation. Lastly, we note that, our analytical framework is not restricted to ESG scores and can be readily extended to other asset-level characteristics, thereby allowing the study of impact investing with respect to alternative dimensions. For readers’ convenience, Table \ref{table intermidiate} summarizes the  optimization problems mentioned above and their corresponding limits of OOS SR.

\begin{table}[h!t]
\centering
\caption{Optimization problems  and the corresponding limit of OOS SRs.}
\label{table intermidiate}
\begin{tabular}{ccc}
\toprule
Optimization Problem                         & Limit of OOS SR      \\ \midrule
$\underset{\bome}{\max}\;\bome\T\hatbmu-\frac{\gamma}{2}\bome\T\left(\hatbSig+\eta   \bQ\right)\bome$  &   $\lim_{T\rightarrow\infty} \frac{ \bmu\T\mathbb{G}\bmu - \frac{\left(\bmu\T\mathbb{G}\tildebs\right)^2}{\tildebs\T\mathbb{G}\tildebs + \frac{1}{T}\Tr\left[\mathbb{G}\bOmega\right]}}{{\sqrt{1-\frac{s_{1,\bSig}}{(1+s_0)^2}}}\sqrt{\bmu\T\mathbb{H}\bmu+\frac{1}{T}\Tr\left[\mathbb{H}\bSig\right]  +
\frac{\left(\bmu\T\mathbb{G}\tildebs\right)^2 \left(\tildebs\T\mathbb{H}\tildebs + \frac{1}{T}\Tr\left[\mathbb{H}\bOmega\right]\right)}{\left(\tildebs\T\mathbb{G}\tildebs + \frac{1}{T}\Tr\left[\mathbb{G}\bOmega\right]\right)^2}  -  2 \frac{\bmu\T\mathbb{G}\tildebs \bmu\T\mathbb{H}\tildebs}{\tildebs\T\mathbb{G}\tildebs + \frac{1}{T}\Tr\left[\mathbb{G}\bOmega\right]}}}$
\\ 
$s.t. \quad \bome^\top \widehat{\tildebs} = 0$ & 
\\
$\underset{\bome}{\max}\;\bome\T\hatbmu-\frac{\gamma}{2}\bome\T\left(\hatbSig+\eta   \bQ\right)\bome$          &  $\lim_{T\rightarrow\infty} \frac{ \bmu\T\mathbb{G}\bmu -\frac{\left(\bmu\T\mathbb{G}\tildebs\right)^2}{\tildebs\T\mathbb{G}\tildebs}}{{\sqrt{1-\frac{s_{1,\bSig}}{(1+s_0)^2}}}\sqrt{\bmu\T\mathbb{H}\bmu+\frac{1}{T}\Tr\left[\mathbb{H}\bSig\right]  +
\frac{\left(\bmu\T\mathbb{G}\tildebs\right)^2 \tildebs\T\mathbb{H}\tildebs}{\left(\tildebs\T\mathbb{G}\tildebs\right)^2}  -  2 \frac{\bmu\T\mathbb{G}\tildebs \bmu\T\mathbb{H}\tildebs}{\tildebs\T\mathbb{G}\tildebs}}}$           \\
 $s.t. \quad\bome^\top {\tildebs} = 0$
&
\\
$\underset{\bome}{\max}\;\bome\T\bmu-\frac{\gamma}{2}\bome\T\left(\hatbSig+\eta   \bQ\right)\bome$                 &  $\lim_{T\rightarrow\infty}\frac{ \bmu\T\mathbb{G}\bmu -\frac{\left(\bmu\T\mathbb{G}\tildebs\right)^2}{\tildebs\T\mathbb{G}\tildebs}}{{\sqrt{1-\frac{s_{1,\bSig}}{(1+s_0)^2}}}\sqrt{\bmu\T\mathbb{H}\bmu  +
\frac{\left(\bmu\T\mathbb{G}\tildebs\right)^2 \tildebs\T\mathbb{H}\tildebs}{\left(\tildebs\T\mathbb{G}\tildebs\right)^2}  -  2 \frac{\bmu\T\mathbb{G}\tildebs \bmu\T\mathbb{H}\tildebs}{\tildebs\T\mathbb{G}\tildebs}}}$   
\\   $s.t. \quad\bome^\top {\tildebs} = 0$
& 
\\
$\underset{\bome}{\max}\;\bome\T\hatbmu-\frac{\gamma}{2}\bome\T\left(\hatbSig+\eta   \bQ\right)\bome$                                        & $\lim_{T\rightarrow\infty}\frac{   \bmu\T\mathbb{G}\bmu}{{\sqrt{1-\frac{s_{1,\bSig}}{(1+s_0)^2}}}\sqrt{\bmu\T\mathbb{H}\bmu+\frac{1}{T}\Tr\left[\mathbb{H}\bSig\right]   }} $                             
\\ \bottomrule      
\end{tabular}
\end{table}



In practice, the selection of the regularization matrix $\eta\bQ$ is important. Different choices  of $\eta \bQ$ may result in different OOS SR performance.
Although Remark \ref{Remark Meng 2025} provides guidance for the optimal choice of the regularization matrix $\bQ$ for the problem in which only the covariance matrix is replaced by its sample counterpart, there still exists a gap between the theoretical maximum Sharpe ratio $\theta_{\max}$ and the OOS SR (shown in Proposition \ref{prop 2}) achieved by the proposed ESG-constrained portfolio. Furthermore,  it is also of interest to determine an optimal tuning parameter $\eta$ that yields the best performance in finite samples given $\bQ$. However, given $\bQ$ and $\eta$, the out of sample Sharpe ratio $\theta^*$  is still not attainable based on historical data since it includes the unknown population parameter $\bmu$ and $\bSig$.  Clearly, if $\theta^{\ast}$ can be consistently estimated using in-sample data, investors can readily evaluate the performance associated with different choices of $\eta\bQ$ and select the one that maximizes the out-of-sample Sharpe ratio of the proposed regularized ESG-constrained portfolio.
Accordingly, we propose an estimator for $\theta^{\ast}$ under a given choice of $\bQ$ and $\eta$.  Lastly, we note that, the estimation of the OOS SR, obtained by simply replacing the unknown $\bmu$ and $\bSig$ with their sample counterparts, is not a good choice  in large-dimensional situations, as the naive empirical estimator often suffers from poor accuracy \citep{ao2019approaching, kan2024sample}.

We define the following quantities:
\begin{align}
\mathcal{D}_{T,1}(\eta,\bQ) &:= {\left(\hatbmu+\widehat{\zeta}\widehat{\tildebs}\right)}^\top\left(\hatbSig + \eta \bQ\right)^{-1} \bmu, \\
\mathcal{D}_{T,2}(\eta,\bQ) &:= {\left(\hatbmu+\widehat{\zeta}\widehat{\tildebs}\right)}^\top\left(\hatbSig + \eta \bQ\right)^{-1} 
{\bSig} 
\left(\hatbSig + \eta {\bQ}\right)^{-1}{\left(\hatbmu+\widehat{\zeta}\widehat{\tildebs}\right)}.
\end{align}
The OOS SR of estimated ESG-constrained portfolio can be re-expressed as
\begin{equation}
\theta^*(\eta,\bQ) = \frac{\mathcal{D}_{T,1}(\eta,\bQ)}{\sqrt{\lvert \mathcal{D}_{T,2}(\eta,\bQ) \rvert}}.
\label{eq oos SR pop}
\end{equation}

\begin{theorem}\label{Theorem SR consistency}
Suppose Assumptions \ref{assum:pT_ratio}--\ref{assum:data_esg} hold true,  define
\begin{align}
&\widehat{\mathcal{D}}_{T,1}(\eta,\bQ):=\hatbmu^\top (\hatbSig + \eta \bQ)^{-1}\hatbmu + \widehat{\zeta}\widehat{\tildebs}^\top   (\hatbSig + \eta \bQ)^{-1}\hatbmu- \frac{\Tr\left((\hatbSig+\eta \bQ)^{-1}\hatbSig\right)}{T-\Tr\left((\hatbSig+\eta \bQ)^{-1}\hatbSig\right)}, \notag\\
&\widehat{\mathcal{D}}_{T,2}(\eta,\bQ):=(\hatbmu+\widehat{\zeta}\widehat{\tildebs})^\top (\hatbSig+\eta \bQ)^{-1}\hatbSig (\hatbSig+\eta \bQ)^{-1}(\hatbmu+\widehat{\zeta}\widehat{\tildebs})/\left(1-\frac{c}{p}\Tr\left(\hatbSig(\hatbSig + \bQ)^{-1}\right)\right)^2.\notag
\end{align}
Then as $T\to\infty$,
\[
\widehat{\mathcal{D}}_{T,1}(\eta,\bQ)-{\mathcal{D}}_{T,1}(\eta,\bQ)\xrightarrow{\mathrm{a.s.}} 0,
\qquad
\widehat{\mathcal{D}}_{T,2}(\eta,\bQ)-{\mathcal{D}}_{T,2}(\eta,\bQ)\xrightarrow{\mathrm{a.s.}} 0.
\]
Accordingly, the estimator of the out-of-sample Sharpe ratio is
\begin{equation}\label{eq SR esti}
\widehat{\theta}^*(\eta,\bQ)
:= \frac{\widehat{\mathcal{D}}_{T,1}(\eta,\bQ)}
{\sqrt{\widehat{\mathcal{D}}_{T,2}(\eta,\bQ)}},
\end{equation}
and satisfies $\widehat{\theta}^*(\eta,\bQ)/{\theta}^*(\eta,\bQ) \overset{a.s}{\longrightarrow}1$.
\end{theorem}

Theorem \ref{Theorem SR consistency} establishes asymptotic consistency of $\widehat{\theta}^*(\eta,\bQ)$ with respect to $\theta^*(\eta,\bQ)$, and thus provides a feasible tool for investors to 
estimate OOS SR of proposed portfolio without requiring the knowledge of $\bmu$ and $\bSig$. Consequently, investors can compare the estimated OOS SR $\widehat{\theta}^\ast$ across different combinations of $(\eta, \bQ)$ and select the one that achieves the highest SR. 
Moreover, with SR estimator $\widehat{\theta}^\ast$ in hand, given a specific $\bQ$, it is natural for investors to select an optimal regularization coefficient $\eta$ by maximizing the OOS SR over a prespecified closed-interval candidate set $\mathcal{G}$, i.e., $$\widehat{\eta}^* = \arg\max_{\eta \in \mathcal{G}} \widehat{\theta}^*(\eta,\bQ).$$ 
As a result, the optimal choice of  $\eta$ for a given  regularization matrix $\bQ$ can be implemented in practice to achieve the best Sharpe ratio performance in finite samples. In our simulation study, we also numerically assess the proximity of $\widehat{\eta}^*$ to the corresponding oracle value $\eta^* := \arg\max_{\eta \in \mathcal{G}} {\theta}^*(\eta,\bQ)$, and the results demonstrate good performance. Finally, we note that, in theory, one could select an optimal $\bQ_\eta :=\eta \bQ$  to maximize the OOS SR estimation; however, since $\bQ_\eta$ is a $p\times p$ positive definite matrix with large, achieving this goal using simple optimization techniques is challenging and time-consuming.

\section{Simulation}
\label{sec: simu}
In this section, we conduct Monte Carlo simulations to assess the performance of the newly developed Sharpe ratio estimator, and to evaluate the impact of different choices of $\bQ$ on the OOS SR.
\subsection{Basic Setup}\label{sec: basic setup}

The virtual return data and ESG score data are both generated from multivariate normal distributions with mean vectors $\bmu_r$ and $\bmu_{esg}$, and covariance matrices $\bSig_r$ and $\bOmega_{esg}$, respectively; that is, $\br_t \sim \mathcal{N}(\bmu_r,\bSig_r), \ba_t \sim \mathcal{N}(\bmu_{esg},\bOmega_{esg})$.
To mimic the real world, the values of population parameters are calibrated from real monthly return and ESG scores data of the largest $p$ components (measured by market values) of the S\&P 500 index from 01/2011 to 12/2019.\footnote{See more data descriptions in the section of empirical analysis.} Specifically,  $\bmu_r$ is  calibrated as the sample means of the asset monthly return, and we set  $\bSig_r$ to be the principal orthogonal
complement thresholding (POET) covariance estimator proposed by \cite{fan2013large} such that the population covariance matrix is positive definite (in the context of $p > T$) and has a factor structure.\footnote{The factor number is set to 2, which is estimated by the selection procedure of \cite{bai2002determining}. The adaptive thresholding estimation of sparse idiosyncratic covariance matrix is applied with thresholding parameter 1 and soft thresholding function \citep{cai2011adaptive, fan2013large}.}
For $\bmu_{esg}$ and $\bOmega_{esg}$, we set them to the sample mean and sample covariance matrix of the ESG scores data.\footnote{For cases where $p > T$, we set $\bOmega_{esg} = \hat{\bOmega}_{esg} + \epsilon_1 I_p$  to ensure positive definiteness, where $\hat{\bOmega}_{esg}$  is the sample covariance matrix estimators and $\epsilon_1$ is set to $10^{-6}$.}

\subsection{Performance of SR estimator}
In this subsection, we evaluate the performance of the newly proposed SR estimator. Specifically, we report the simulation results for SR estimation shown in \eqref{eq SR esti} and the corresponding oracle OOS SR shown in \eqref{eq SR}. For our regularized ESG-constraint portfolio,  we consider two specifications for the regularization matrix $\bQ$. (1) The first sets $\bQ = \bSig_r$, corresponding to the population covariance matrix, which is theoretically optimal when only sample covariance matrix is applied, see Remark \ref{Remark Meng 2025}. (2) The second sets $\bQ = I_p$, the identity matrix, which serves as a naive benchmark. Clearly, $I_p$ only regularizes the eigenvalues, whereas $\bSig_r$ not only adjusts the eigenvalues but also incorporates additional structural information. We note that, when $\bQ = I_p$, $\hatbSig+\eta \bQ$ is equivalent to the shrinkage covariance estimator proposed by \cite{ledoit2004well} with shrinking intensity $\eta$.
For the tuning parameter $\eta$, its values are set to range from 0.2 to 10 with an increment of 0.2 in the first case, and from 0.0002 to 0.01 with an increment of 0.0002 in the second case. 
The expected portfolio ESG score $\bars$ is set to 0.8, which is larger than the ESG score of unconstrained oracle portfolio. 
For the combinations of $p$ and $T$, we consider $(p, T) \in {(60, 120), (180, 120), (180, 360)}$, which correspond to $p/T$ ratios of 0.5 and 1.5. 
The simulation is replicated 1,000 times, and the reported results are the averages across these replications.

\begin{figure}[!htb]
\centering	
\subfloat[Situation when $Q = \bSig_r$]
{\includegraphics[width=7.7cm]{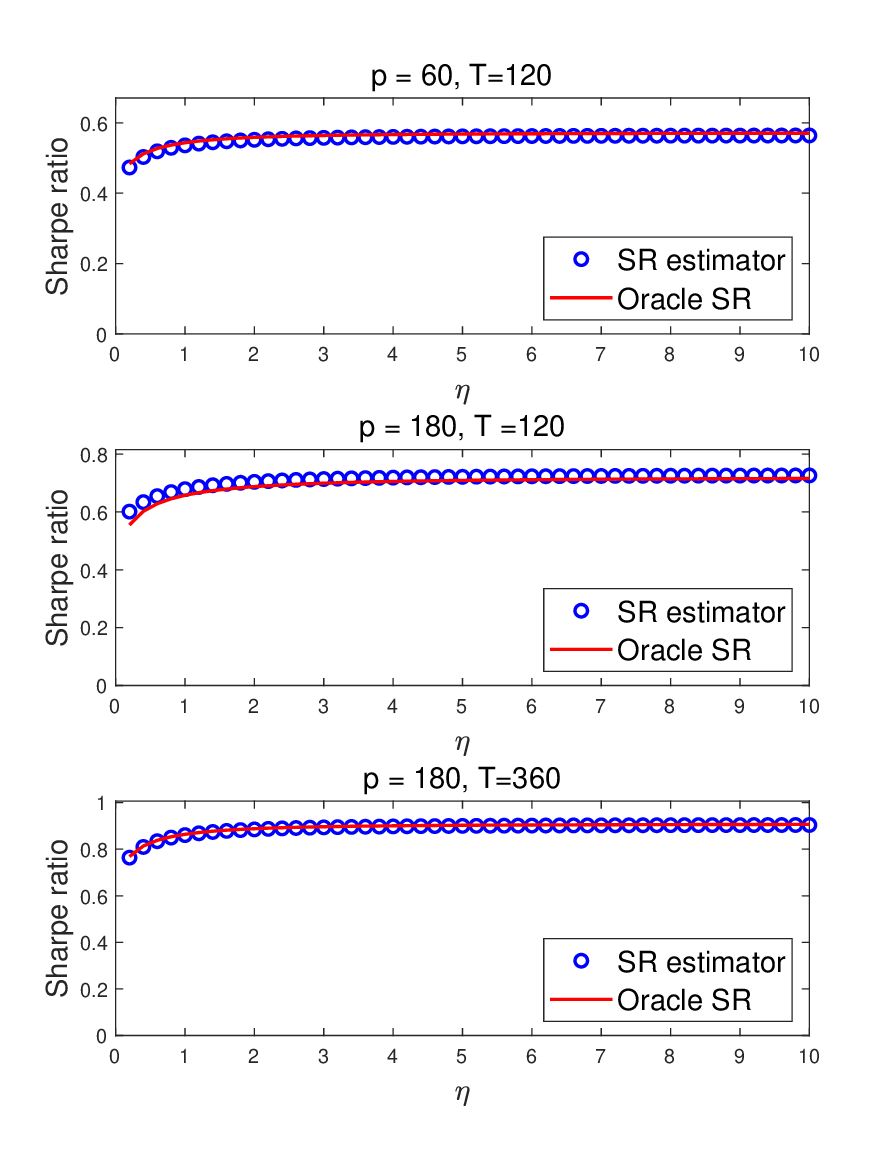}\label{Figure SR Q=Error}}
\quad
\subfloat[Situation when $Q = I_p$]
{\includegraphics[width=7.7cm]{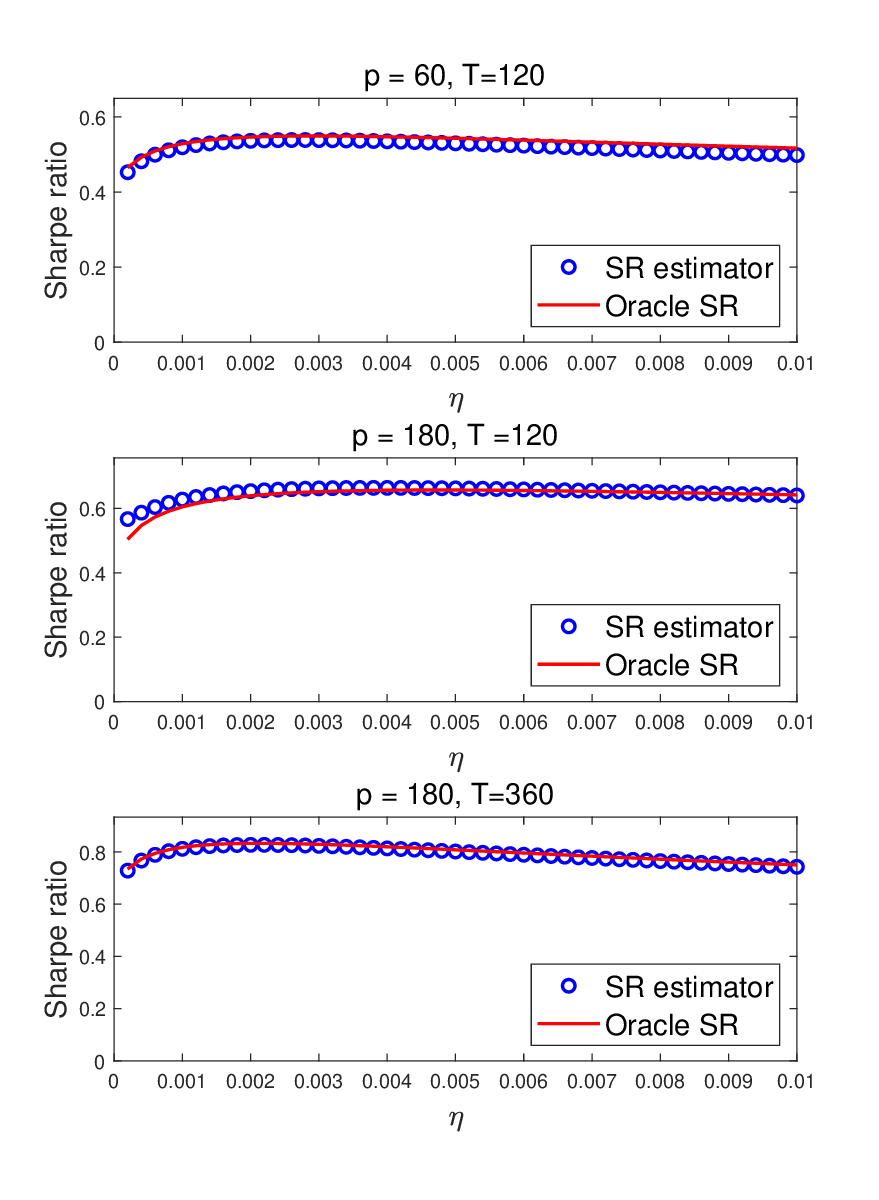}\label{Figure SR Q=I}}
\caption{Simulation results of Sharpe ratio estimator and the corresponding oracle level with respect to $\eta$.}
\end{figure}

Figure \eqref{Figure SR Q=Error} and \eqref{Figure SR Q=I} show the simulation results for both situations. First of all, it is evident that the value of $\widehat{\theta}^*(\eta,\bQ)$ matches the value of ${\theta}^*(\eta,\bQ)$  well across various $\eta$ in most cases.  As sample size $T$ increases, the SR estimations $\widehat{\theta}^*(\eta,\bQ)$ becomes closer to the corresponding oracle level. Second,  when $\bQ = \bSig_r$, we can observe that, as $\eta$ increases, the OOS SR of the proposed ESG-constrained portfolio increases. This is because  the matrix $\hat{\bSig} + \eta \bSig_r$ becomes increasingly dominated by the term $\eta \bSig_r$, which is proportional to the true covariance matrix. In addition,  
when $\bQ = \bI_p$, it can be observed that ${\theta}^*(\eta,\bQ)$ first increases and then decreases as $\eta$ grows, indicating the existence of an optimal $\eta$ that maximizes the OOS SR. More importantly, we can observe that the proposed SR estimator $\widehat{\theta}^\ast$ closely replicates the pattern of $\theta^\ast$. Consequently, in practical applications, one can select the value of $\eta$ that maximizes the OOS SR without requiring knowledge of the true population parameters, and thus determine the optimal regularization matrix $\eta\bQ$ for a given $\bQ$ over a deterministic candidate set $\mathcal{G}$ of $\eta$.

\begin{figure}[!htb]
\centering	
\subfloat[Situation for $p/T = 0.5$]
{\includegraphics[width=7.7cm]{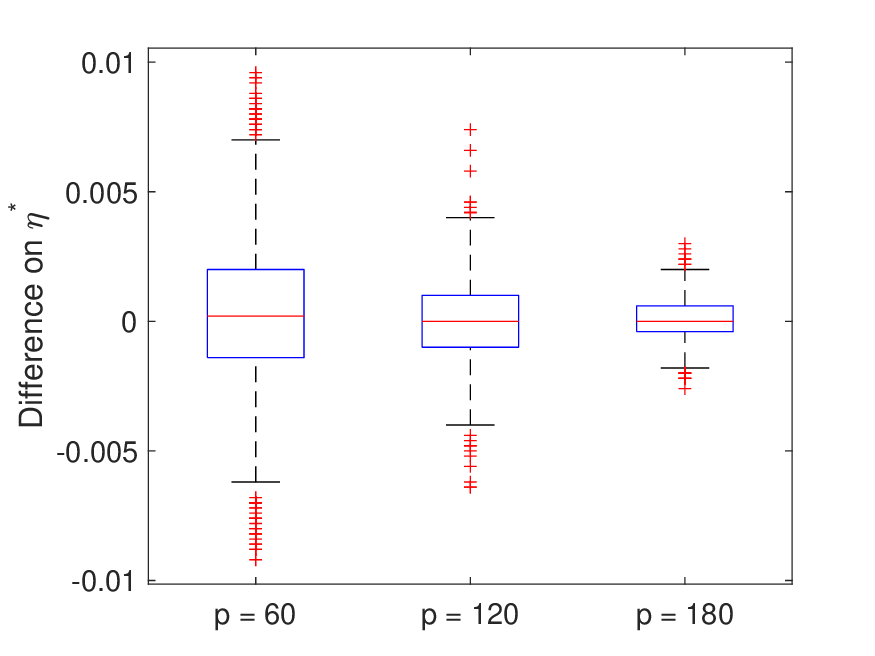}\label{Figure eta fix c}}
\quad
\subfloat[Situation for varying $p/T$ ratio, $p = 180$]
{\includegraphics[width=7.7cm]{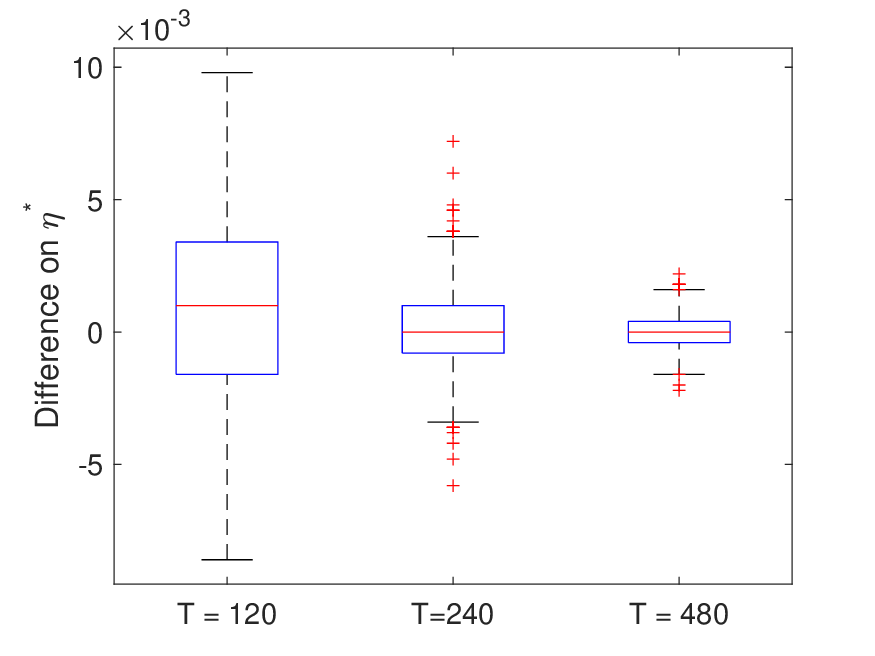}\label{Figure eta increase T}}
\caption{Boxplots of the difference between estimated Sharpe ratio  and the corresponding oracle level.}
\end{figure}

Next, we further evaluate the accuracy of the optimal $\hat{\eta}^*$ selected based on  $\widehat{\theta}^{\ast}$, by examining the discrepancy between $\widehat{\eta}^{\ast}$ and the population level $\eta^{\ast}$ that maximizes the true (but unknown) $\theta^{\ast}$, i.e., $\eta^* = \argmax_{\eta\in \mathcal{G}} \theta^*(\eta,\bQ)$. We focus on the case where $\bQ = \bI_p$, since Figure~\eqref{Figure SR Q=I} indicates that the optimal $\eta^{\ast}$ lies in the interior of the candidate set $\mathcal{G} = [0.0002,0.01]$.

Specifically, the y-axis of Figures \eqref{Figure eta fix c} and \eqref{Figure eta increase T} gives the boxplot of the discrepancies (i.e., $\argmax_{\eta\in\mathcal{G}
} \theta^*(\eta,\bQ) - \argmax_{\eta\in\mathcal{G}} \widehat{\theta}^*(\eta,\bQ)$) over 1000 independent simulation experiments for various situations. In Figure \eqref{Figure eta fix c}, the $p/T$ ratio is fixed at 0.5 with $p = 60, 120,$ and $180$. In Figure \eqref{Figure eta increase T}, we fix $p = 180$ and consider increasing sample sizes of $T = 120, 240,$ and $480$ such that $p/T$ ratio decreases.
It can be observed that, under a fixed $p/T$ ratio, as the sample size increases, the variability of the difference between  the optimal $\eta^*$ and its estimator $\widehat{\eta}^*$ decreases, and its mean value becomes closer to zero. Similarly, as the ratio $p/T$ decreases (i.e., with a larger sample size), the discrepancy between $\widehat{\eta}^{\ast}$ and $\eta^{\ast}$ diminishes. As such, these results indicate that selecting the optimal value of $\eta$ based on the proposed Sharpe ratio estimator is approximately as efficient as using the true Sharpe ratio $\theta^{\ast}$, highlighting the excellent performance of our out-of-sample Sharpe ratio estimator.

\subsection{The Choices of Regularization Matrix}
In this subsection, we evaluate the performance of the proposed regularized portfolios by examining how different choices of $\bQ$ affect the out-of-sample financial and sustainable performance of the portfolio. Specifically, we consider the OOS average return $AM$, standard deviation $SD$, Sharpe ratio $SR$, average portfolio ESG score $AM_{esg}$, and the standard deviation of portfolio ESG scores $SD_{esg}$ for estimated portfolio, which are defined respectively as follows: 
\begin{align*}
    &AM = \widehat{\bome}^\top\bmu, \quad SD = \sqrt{\widehat{\bome}^\top\bSig \widehat{\bome}}, \quad SR = \frac{\widehat{\bome}^\top\bmu}{\sqrt{\widehat{\bome}^\top\bSig \widehat{\bome}}},
    \\ & AM_{esg} = \widehat{\bome}^\top \mu_{esg}, \quad SD_{esg} = \sqrt{\widehat{\bome}^\top\bOmega_{esg} \widehat{\bome}}.
\end{align*}
Here, $\widehat{\bome}$ denotes the scaled estimated portfolio such that $\widehat{\bome}^\top \bm{1} = 1$.

 Since we are interested in the choice of regularization term $\eta\bQ$, we consider the following several specifications for $\bQ$: (1) Identity matrix, denoted as $\bI_p$;
(2) Diagonal matrix of sample return covariance matrix, denoted as $\hat{\bSig}_{d}$;
(3) Linear shrinkage covariance matrix estimation proposed by \cite{ledoit2004well}, denoted as $\hatbSig_{linear}$, the shrinkage intensity is determined following \cite{ledoit2003improved};
(4) Nonlinear shrinkage covariance matrix estimation proposed by \cite{ledoit2017nonlinear}, denoted as $\hatbSig_{Nonlinear}$;
(5) POET covariance estimation proposed by \cite{fan2013large}, $\hatbSig_{POET}$, the factor number is selected by the procedure of \cite{bai2002determining}, and the sparse covariance matrix is estimated by using soft thresholding function with threshold parameter 0.5; 
(6) Sample covariance matrix of assets' ESG scores, denoted as $\widehat{\bOmega}_{esg}$, and we use  $\widehat{\bOmega}_{esg} + 10^{-6}I_p$ to ensure the positive definiteness when $p > T$; (7)  Diagonal matrix whose diagonal elements are the reciprocals of the corresponding sample mean ESG scores, denoted as $\bD^{-1}$;
(8) Diagonal matrix whose diagonal elements are the corresponding sample mean ESG scores, denoted as $\bD$;
(9) Population return covariance matrix, denoted as $\bSig_{r}$; (10) Population covariance matrix of assets' ESG scores, denoted as $\bOmega_{esg}$.
Case (1) is a naive choice that ensures the positive definiteness of $\widehat{\bSig}+\eta\bQ$; Cases (2)–(5) employ return covariance matrix estimators commonly used in the portfolio allocation literature as the regularization matrix. Cases (6)–(8) incorporate information from ESG score variables for penalization. Cases (9) and (10) rely on unknown population quantities, which are infeasible in practice and therefore serve only as benchmark cases.
 We report the results using $\widehat{\eta}^*$ and $\eta^*$. The results of our new portfolio are labeled as Re-MV (\textbf{R}egularized \textbf{e}sg-constrained \textbf{M}ean-\textbf{V}ariance portfolio). The candidate set $\mathcal{G}$ ranges from 0.0002 to 0.01 with an increment of 0.0002 for Case (1), from 0.02 to 1 with an increment of 0.02 for Cases (7)–(8), and from 0.2 to 10 with an increment of 0.2 for the remaining cases.

For comparison, the results of following benchmarks are reported: (1) Oracle: mean-variance portfolio using population parameter values in \eqref{eq classic problem}; (2) Sample: mean-variance portfolio using sample mean and sample covariance matrix in \eqref{eq classic problem}; (3) M-MV-O: mean-variance portfolio with ESG level constraint shown in \eqref{eq optimization ESG level only} using population parameters; (4) M-MV-S: mean-variance portfolio with ESG level constraint shown in \eqref{eq optimization ESG level only} using sample mean return, sample covariance of return, and sample mean ESG score; 
(5) Q-MV: regularized mean-variance portfolio in \eqref{eq: empirical version 3} without ESG level constraint using sample mean and sample covariance.

\begin{table}[h!t]
\centering
\caption{Simulation results for considered strategies across various regularization matrix $\bQ$. $p =180, T =360$, respectively. The target ESG level $\bars$ is set to $0.8$.}
\label{Table simu Q1}
\tabcolsep 0.056in
\begin{tabular}{lcccccccccc}
 \toprule & \multicolumn{1}{l}{$AM$} & \multicolumn{1}{l}{$SD$} & \multicolumn{1}{l}{$SR$} & $AM_{esg}$ & $SD_{esg}$                 & \multicolumn{1}{l}{$AM$} & \multicolumn{1}{l}{$SD$} & \multicolumn{1}{l}{$SR$} & $AM_{esg}$ & $SD_{esg}$ \\ \midrule
Oracle                & 0.0850                   & 0.074                    & 1.154                    & -0.402     & 0.268                      & -                        & -                        & -                        & -          & -          \\
M-MV-O                & 0.1097                   & 0.101                    & 1.083                    & 0.800      & 0.234                      & -                        & -                        & -                        & -          & -          \\
Sample                & 0.1403                   & 0.205                    & 0.697                    & -1.139     & 0.534                      & -                        & -                        & -                        & -          & -          \\
M-MV-S                & 0.1476                   & 0.630                    & 0.592                    & 0.778      & 1.373                      & -                        & -                        & -                        & -          & -          \\
\midrule
& \multicolumn{5}{c}{$\bQ = \bI_p$}                                                                                          & \multicolumn{5}{c}{$\bQ = \hatbSig_d$}                                                                   \\
Q-MV($\hat{\eta}^*$)  & 0.0498                   & 0.056                    & 0.897                    & 0.153      & \multicolumn{1}{c|}{0.145} & 0.0479                   & 0.052                    & 0.925                    & 0.145      & 0.150      \\
Q-MV(${\eta}^*$)      & 0.0491                   & 0.055                    & 0.901                    & 0.162      & \multicolumn{1}{c|}{0.143} & 0.0473                   & 0.051                    & 0.928                    & 0.153      & 0.148      \\
Re-MV($\hat{\eta}^*$) & 0.0615                   & 0.074                    & 0.830                    & 0.799      & \multicolumn{1}{c|}{0.133} & 0.0562                   & 0.066                    & 0.856                    & 0.799      & 0.121      \\
Re-MV(${\eta}^*$)     & 0.0602                   & 0.072                    & 0.834                    & 0.799      & \multicolumn{1}{c|}{0.129} & 0.0551                   & 0.064                    & 0.860                    & 0.799      & 0.118      \\ \midrule
& \multicolumn{5}{c}{$\bQ = \bD$}                                                                                          & \multicolumn{5}{c}{$\bQ = \bD^{-1}$}                                                                     \\
Q-MV($\hat{\eta}^*$)  & 0.0286 & 0.037 & 0.777 & 0.385 & \multicolumn{1}{c|}{0.092} & 0.0214 & 0.034 & 0.632 & 0.569 & 0.055 \\
Q-MV(${\eta}^*$)      & 0.0286 & 0.037 & 0.777 & 0.385 & \multicolumn{1}{c|}{0.092} & 0.0214 & 0.034 & 0.632 & 0.569 & 0.055 \\
Re-MV($\hat{\eta}^*$) & 0.0339 & 0.046 & 0.729 & 0.799 & \multicolumn{1}{c|}{0.075} & 0.0236 & 0.038 & 0.613 & 0.800 & 0.037 \\
Re-MV(${\eta}^*$)     & 0.0339 & 0.046 & 0.729 & 0.799 & \multicolumn{1}{c|}{0.075} & 0.0236 & 0.038 & 0.613 & 0.800 & 0.037      \\ \midrule
                      & \multicolumn{5}{c}{$\bQ = \hatbSig_{linear}$}                                                                            & \multicolumn{5}{c}{$\bQ = \hatbSig_{Nonlinear}$}                                                         \\
Q-MV($\hat{\eta}^*$)  & 0.0951                   & 0.126                    & 0.756                    & -0.508     & \multicolumn{1}{c|}{0.333} & 0.067                    & 0.075                    & 0.900                    & -0.078     & 0.196      \\
Q-MV(${\eta}^*$)      & 0.0951                   & 0.126                    & 0.756                    & -0.508     & \multicolumn{1}{c|}{0.333} & 0.067                    & 0.075                    & 0.900                    & -0.079     & 0.196      \\
Re-MV($\hat{\eta}^*$) & 0.2392                   & 0.418                    & 0.683                    & 0.781      & \multicolumn{1}{c|}{0.865} & 0.096                    & 0.116                    & 0.831                    & 0.799      & 0.215      \\
Re-MV(${\eta}^*$)     & 0.2392                   & 0.418                    & 0.683                    & 0.781      & \multicolumn{1}{c|}{0.865} & 0.097                    & 0.116                    & 0.831                    & 0.799      & 0.215      \\ \midrule
  & \multicolumn{5}{c}{$\bQ = \widehat{\bOmega}_{esg}$}                                                                      & \multicolumn{5}{c}{$\bQ = {\bOmega}_{esg}$}                                                              \\
Q-MV($\hat{\eta}^*$)  & 0.0839                   & 0.120                    & 0.702                    & -0.263     & \multicolumn{1}{c|}{0.005} & 0.0838                   & 0.120                    & 0.703                    & -0.261     & 0.004      \\
Q-MV(${\eta}^*$)      & 0.0834                   & 0.119                    & 0.703                    & -0.257     & \multicolumn{1}{c|}{0.003} & 0.0834                   & 0.119                    & 0.703                    & -0.257     & 0.003      \\
Re-MV($\hat{\eta}^*$) & 0.0743                   & 0.277                    & 0.646                    & 0.800      & \multicolumn{1}{c|}{0.011} & -2.0725                  & 3.718                    & 0.646                    & 0.801      & 0.092      \\
Re-MV(${\eta}^*$)     & 0.0684                   & 0.267                    & 0.646                    & 0.800      & \multicolumn{1}{c|}{0.007} & -2.0767                  & 3.711                    & 0.647                    & 0.801      & 0.089      \\ \midrule
                      & \multicolumn{5}{c}{$\bQ = \hatbSig_{POET}$}                                                                              & \multicolumn{5}{c}{$\bQ = \bSig_r$}                                                                      \\
Q-MV($\hat{\eta}^*$)  & 0.1232                   & 0.282                    & 0.844                    & -0.769     & \multicolumn{1}{c|}{0.854} & 0.089                    & 0.091                    & 0.982                    & -0.455     & 0.290      \\
Q-MV(${\eta}^*$)      & 0.1274                   & 0.408                    & 0.827                    & -1.010     & \multicolumn{1}{c|}{1.226} & 0.089                    & 0.090                    & 0.983                    & -0.455     & 0.290      \\
Re-MV($\hat{\eta}^*$) & 0.2234                   & 0.775                    & 0.623                    & 0.806      & \multicolumn{1}{c|}{1.654} & 0.123                    & 0.136                    & 0.906                    & 0.798      & 0.300      \\
Re-MV(${\eta}^*$)     & 0.6013                   & 1.594                    & 0.479                    & 0.794      & \multicolumn{1}{c|}{3.239} & 0.123                    & 0.136                    & 0.907                    & 0.798      & 0.300     \\ \bottomrule
\end{tabular}
\end{table}

For the optimization with the ESG-level constraint, we again set the threshold to 0.8,\footnote{The ESG score of oracle mean-variance portfolio is less than 0.8 in our simulation.}  and focus on simulation replications where the out-of-sample ESG score of the sample mean–variance portfolio falls below this value.\footnote{When the ESG score of the sample mean–variance portfolio exceeds the preset threshold $\bars$, the ESG-level constraint becomes non-binding for M-MV-S and Re-MV, and thus M-MV-S degenerates to sample and Re-MV degenerates to Q-MV.}  The results of our method is labeled as Re-MV. Two  combinations  $\{p=180, T=120\}$ and $\{p=180, T=360\}$ of $p$ and $T$ are considered, risk aversion parameter $\gamma$ is set to 5, and simulation results are the average level over 1000 replications.

\begin{table}[h!t]
\centering
\caption{Simulation results for considered strategies across various regularization matrix $\bQ$. $p =180, T =120$, respectively. The target ESG level $\bars$ is set to $0.8$.}
\label{Table simu Q2}
\tabcolsep 0.056in
\begin{tabular}{lcccccccccc}
\toprule 
  & \multicolumn{1}{l}{$AM$} & \multicolumn{1}{l}{$SD$} & \multicolumn{1}{l}{$SR$} & $AM_{esg}$ & $SD_{esg}$ & \multicolumn{1}{l}{$AM$} & \multicolumn{1}{l}{$SD$} & \multicolumn{1}{l}{$SR$} & $AM_{esg}$ & $SD_{esg}$ \\ \midrule
Oracle                & 0.0850                   & 0.074                    & 1.154                    & -0.402     & 0.268                      & -                        & -                        & -                        & -          & -          \\
M-MV-O                & 0.1097                   & 0.101                    & 1.083                    & 0.800      & 0.234                      & -                        & -                        & -                        & -          & -          \\ \midrule
                      & \multicolumn{5}{c}{$\bQ = \bI_p$}                                                                                          & \multicolumn{5}{c}{$\bQ = \hatbSig_d$}                                                                   \\
Q-MV($\hat{\eta}^*$)  & 0.0470                   & 0.072                    & 0.689                    & 0.220      & \multicolumn{1}{c|}{0.152} & 0.0415                   & 0.056                    & 0.744                    & 0.250      & 0.139      \\
Q-MV(${\eta}^*$)      & 0.0357                   & 0.049                    & 0.727                    & 0.350      & \multicolumn{1}{c|}{0.106} & 0.0350                   & 0.046                    & 0.764                    & 0.329      & 0.113      \\
Re-MV($\hat{\eta}^*$) & 0.0636                   & 0.157                    & 0.618                    & 0.801      & \multicolumn{1}{c|}{0.239} & 0.0502                   & 0.081                    & 0.675                    & 0.798      & 0.140      \\
Re-MV(${\eta}^*$)     & 0.0436                   & 0.066                    & 0.661                    & 0.798      & \multicolumn{1}{c|}{0.100} & 0.0402                   & 0.057                    & 0.699                    & 0.798      & 0.093      \\ \midrule
                      & \multicolumn{5}{c}{$\bQ = \bD$}                                                                                          & \multicolumn{5}{c}{$\bQ = \bD^{-1}$}                                                                     \\
Q-MV($\hat{\eta}^*$)  & 0.0277                   & 0.040                    & 0.691                    & 0.397      & \multicolumn{1}{c|}{0.091} & 0.0214                   & 0.036                    & 0.600                    & 0.570      & 0.056      \\
Q-MV(${\eta}^*$)      & 0.0277                   & 0.040                    & 0.691                    & 0.398      & \multicolumn{1}{c|}{0.091} & 0.0214                   & 0.036                    & 0.600                    & 0.570      & 0.056      \\
Re-MV($\hat{\eta}^*$) & 0.0338                   & 0.054                    & 0.627                    & 0.798      & \multicolumn{1}{c|}{0.082} & 0.0239                   & 0.042                    & 0.569                    & 0.799      & 0.040      \\
Re-MV(${\eta}^*$)     & 0.0337                   & 0.054                    & 0.627                    & 0.798      & \multicolumn{1}{c|}{0.081} & 0.0239                   & 0.042                    & 0.569                    & 0.799      & 0.040      \\ \midrule
                      & \multicolumn{5}{c}{$\bQ = \hatbSig_{linear}$}                                                                            & \multicolumn{5}{c}{$\bQ = \hatbSig_{Nonlinear}$}                                                         \\
Q-MV($\hat{\eta}^*$)  & 0.062                    & 0.180                    & 0.559                    & -0.094     & \multicolumn{1}{c|}{0.366} & 0.051                    & 0.072                    & 0.722                    & 0.158      & 0.152      \\
Q-MV(${\eta}^*$)      & 0.062                    & 0.180                    & 0.559                    & -0.094     & \multicolumn{1}{c|}{0.366} & 0.052                    & 0.072                    & 0.722                    & 0.153      & 0.154      \\
Re-MV($\hat{\eta}^*$) & 0.062                    & 0.458                    & 0.456                    & 0.756      & \multicolumn{1}{c|}{0.787} & 0.062                    & 0.129                    & 0.640                    & 0.800      & 0.207      \\
Re-MV(${\eta}^*$)     & 0.062                    & 0.458                    & 0.456                    & 0.756      & \multicolumn{1}{c|}{0.787} & 0.064                    & 0.131                    & 0.641                    & 0.800      & 0.211      \\ \midrule
                      & \multicolumn{5}{c}{$\bQ = \widehat{\bOmega}_{esg}$}                                                                      & \multicolumn{5}{c}{$\bQ = {\bOmega}_{esg}$}                                                              \\
Q-MV($\hat{\eta}^*$)  & 0.0652                   & 0.363                    & 0.367                    & -0.119     & \multicolumn{1}{c|}{0.008} & 0.030                    & 0.503                    & 0.365                    & 0.129      & 0.011      \\
Q-MV(${\eta}^*$)      & 0.1358                   & 0.391                    & 0.477                    & -0.848     & \multicolumn{1}{c|}{0.008} & 0.064                    & 0.369                    & 0.381                    & -0.082     & 0.008      \\
Re-MV($\hat{\eta}^*$) & -0.0406                  & 0.632                    & 0.280                    & 0.800      & \multicolumn{1}{c|}{0.014} & 0.259                    & 1.223                    & 0.281                    & 0.802      & 0.026      \\
Re-MV(${\eta}^*$)     & 0.0702                   & 0.430                    & 0.384                    & 0.800      & \multicolumn{1}{c|}{0.009} & 0.065                    & 0.980                    & 0.295                    & 0.800      & 0.021      \\ \midrule
                      & \multicolumn{5}{c}{$\bQ = \hatbSig_{POET}$}                                                                              & \multicolumn{5}{c}{$\bQ = \bSig_r$}                                                                      \\
Q-MV($\hat{\eta}^*$)  & 0.166                    & 0.436                    & 0.526                    & -1.332     & \multicolumn{1}{c|}{1.037} & 0.0887                   & 0.177                    & 0.752                    & -0.468     & 0.471      \\
Q-MV(${\eta}^*$)      & -0.059                   & 1.075                    & 0.521                    & 2.331      & \multicolumn{1}{c|}{2.443} & 0.1048                   & 0.147                    & 0.780                    & -0.653     & 0.410      \\
Re-MV($\hat{\eta}^*$) & 0.844                    & 1.790                    & 0.363                    & 1.027      & \multicolumn{1}{c|}{4.957} & 0.0470                   & 0.404                    & 0.634                    & 0.812      & 0.839      \\
Re-MV(${\eta}^*$)     & -0.048                   & 1.005                    & 0.278                    & 0.788      & \multicolumn{1}{c|}{1.904} & 0.1321                   & 0.427                    & 0.663                    & 0.800      & 0.923    \\ \bottomrule  
\end{tabular}
\end{table}

Tables \ref{Table simu Q1}-\ref{Table simu Q2} report the simulation results for various forms of $\bQ$ under considered two sample-dimension combinations, respectively.  Several findings can be summarized as follows: (1) First, the two oracle portfolios achieve the highest SRs, 1.154 and 1.083, respectively. The unconstrained oracle portfolio exhibits a low ESG score of $-0.402$, whereas the ESG-constrained oracle portfolio satisfies the required ESG level. Imposing an ESG-level constraint can substantially improve OOS ESG performance with only a slight sacrifice in portfolio SR. (2) Second, compared to the oracle portfolios, the constrained and unconstrained sample mean–variance portfolios achieve much lower Sharpe ratios, 0.697 and 0.592, respectively, and the unconstrained sample portfolio exhibits a substantially lower ESG level of $-1.139$ (in the case of $p<T$). (3) Third, for the regularized (un)constrained portfolios, it is evident that different choices of the regularization matrix $\bQ$ lead to different performance outcomes. Specifically, when $\bQ=\bSig_r$, both Q-MV and Re-MV achieve the highest OOS SRs. However, $\bSig_r$ is unattainable in practice. Among all feasible choices of $\bQ$, setting $\bQ=\hatbSig_d$ yields the highest OOS SRs for the regularized portfolios,  substantially improving on the performance of the sample portfolio. Furthermore, the corresponding ESG level of the Re-MV portfolio  is very close to the required standard  0.8. In the case of $p>T$, the OOS SRs of Re-MV with $\bQ=\widehat{\bSig}_d$ using $\hat{\eta}^\ast$ and $\eta^\ast$ are 0.675 and 0.699, respectively, both of which exceed the corresponding values of 0.634 and 0.663 obtained by Re-MV with $\bQ=\bSig_r$ using $\hat{\eta}^\ast$ and $\eta^\ast$, respectively.
The second-tier SR performance is achieved by choosing $\bQ=I_p$ and $\bQ=\hatbSig_{Nonlinear}$. We note that, since the population parameters are calibrated using real return and ESG score data, we believe that these choices are capable of delivering good performance in practical applications and can therefore serve as useful guidance for practical implementation. (4) Under the situations in which $\bQ$ is set to the (estimated) covariance of  ESG scores (i.e. $\bQ = \bOmega_{esg},\widehat{\bOmega}_{esg}$), the regularized portfolios exhibit SRs similar to those of the corresponding sample portfolios and provide slight improvement in SR performance. Based on the economic interpretation in Remark \ref{remark economic}, for an ESG-concerned investor who solves the mean–variance optimization problem with additional ESG-mean and ESG-variance constraints (i.e., problem \eqref{eq ESG problem}), imposing an ESG risk constraint to restrict the portfolio ESG variance contributes little to improving SR performance. As a result, for data providers, ESG measure designers, and even policymakers, adjusting the covariance structure of assets’ ESG scores to align with other considered forms of $\bQ$ (e.g., $\hatbSig_d$) can enable ESG-motivated investors to achieve better financial performance in high-dimensional settings when portfolio ESG-variance constraints are further imposed. (5) Comparing the results obtained with $\bQ=\bD$ and $\bQ=\bD^{-1}$, it is evident that using ESG score levels yields better  SR performance than using the reciprocals of ESG levels, as considered in \cite{makridis2023balancing}. 
 (6) Next, it can be observed that the performances of Q-MV and Re-MV based on $\widehat{\theta}^*$ are close to those obtained using the oracle ${\theta}^*$, indicating that the estimation of the optimal $\eta$ is accurate and performs well in practice. (7) It is evident that, with an appropriately chosen $\bQ$, the newly proposed optimization problem performs well in the case of $p>T$. As such, our method extends the classic sample portfolio to settings where $p>T$.
  (8) Moreover, based on the ESG variance measure $SD_{esg}$, it is evident that Re-MV also achieves the lower OOS portfolio ESG volatility, compared to both oracle portfolios and sample portfolios. Although choosing $\bQ$ different from $\bOmega_{esg}$  does not explicitly restrict the variance of the portfolio ESG scores, it still leads to improved stability of ESG outcomes. (9) Finally, comparing the results of Q-MV and Re-MV reveals that, in high-dimensional settings, the desired sustainable ESG level can be attained by imposing the ESG-level constraint, without incurring a significant reduction in the SR. As a consequence, with an appropriately chosen regularization matrix $\bQ$, the proposed portfolio can achieve a stable ESG score that satisfies the required level while maintaining an ideal OOS SR.

\section{Empirical application}
\label{sec: empirical}
In this section, we evaluate the out-of-sample economic and ESG performance of our proposed regularized ESG-constrained mean-variance portfolio using real market data. Specifically, the evaluation procedure is conducted under the rolling-window scheme. At each decision point, we use the historical data of length $T$ to construct portfolios, which are then held for the next several periods (say $T_{hold}$) to gain the OOS realized excess portfolio return and OOS portfolio ESG score. After holding portfolios for $T_{hold}$ periods, we re-balance the portfolio allocation using the updated data till the new decision point. This process is repeated until the end of the out-of-sample periods.

\subsection{Data}
We consider monthly return data of  component stocks of  the S\&P 500 index, sourced from the Center for Research in Security Prices (CRSP). The S\&P 500 index consists of large-cap stocks, representing the largest publicly traded companies in the U.S. The data cover the period from January 2010 to December 2024, comprising 180 monthly observations over 15 years. We delete those stocks that do not have complete historical data at the first decision point. To avoid look-ahead bias due to missing data during the rolling window process, we simply drop the corresponding assets from the basket \citep{ao2019approaching,fan2022,wu2025making}.
The risk-free data are downloaded from the Ken French's data library.\footnote{\url{https://mba.tuck.dartmouth.edu/pages/faculty/ken.french/data_library.html}} For the ESG score data corresponding to each asset in the considered investment universe over the considered periods, we obtained them from the Refinitiv ESG dataset provided by London Stock Exchange Group (LSEG).\footnote{\url{https://www.lseg.com/en}} LSEG ESG provides one of the most comprehensive ESG datasets in the industry, covering over 88\% of the global market capitalization and more than 700 individual ESG metrics.

Based on the components of the S\&P 500 index, we consider two scenarios:
(1) the largest (randomly selected) $p = 60$ stocks (measured by market capitalization) with a sample size of $T = 90$, thus the sample size is larger than the number of assets; and
(2) the largest (randomly selected) $p = 180$ stocks with a sample size of $T = 90$, thus the sample size is smaller than the dimension. The out of sample periods for both cases is from 06/2017 to 12/2024, also covering 90 periods. 

\subsection{Portfolios and Measures}
For the newly proposed regularized ESG-constrained portfolio, our simulation results indicate the importance of selecting an appropriate form of the regularization matrix $\bQ$, as it has a substantial impact on portfolio SR performance. Although our simulations suggest that $\bQ=\hatbSig_d$ is the best feasible choice among considered candidates, practical settings are likely to be considerably more complex than the simulation environment considered here.
Fortunately, the availability of proposed OOS SR estimator  $\widehat{\theta}^*$ allows us to evaluate portfolio OOS performance at each decision node based on historical data. Therefore, we can estimate the OOS SRs for portfolios constructed with different choices of $\bQ$ and the corresponding estimated $\hat{\eta}^\ast$, and  select the regularization matrix $\hat{\eta}^\ast \bQ$ that yields the highest estimated SR. In our empirical studies, we apply this adaptive strategy and consider $\bQ = \hatbSig_d$, $ \hatbSig_{Nonlinear}$, and $I_p$ as candidate choices. The results are labeled as Re-MV. The candidate set $\mathcal{G}$ for $\eta$ takes the same range as those used in our simulation study.

As benchmarks, beyond the feasible strategies introduced in the simulation section, we further include the following six portfolios for comparison: (1)-(2) mean–variance portfolio with (without) ESG level constraint (M-MV-POET / MV-POET), based on factor-model-estimated means and the POET covariance estimator \citep{fan2013large, wu2025making}; The number of common factors is selected by BIC criterion of \cite{bai2002determining}, and the adaptive sparse  thresholding parameter is set to 0.5. (3)-(4) Mean-variance portfolio with (without) ESG level constraint (M-MV-Li / MV-Li), based on sample mean and linear shrinkage estimator of \cite{ledoit2004well}; (3) mean-variance portfolio with (without) ESG level constraint (M-MV-NL / MV-NL), based on sample mean and nonlinear shrinkage estimator of \cite{ledoit2017nonlinear}. For the Q-MV strategy, we apply the same procedure for selecting $\hat{\eta}^*\bQ$ as that used for Re-MV.
For portfolios with ESG level constraint, we set the target ESG score to 0.8 and 0.9. The holding periods for all portfolios are set to one month, corresponding to $T_{hold} = 1$.

 For evaluating measures, we report the results for OOS SR, OOS average ESG score, the 25$^{th}$ percentile of OOS ESG scores, and the standard deviation of OOS ESG score, defined as follows, respectively:
\begin{align*}
& \text{SR} = \frac{\frac{1}{\mathbb{T}}\sum_{t=1}^{\mathbb{T}} r_{p,t}}{ \sqrt{\frac{1}{\mathbb{T}-1}\sum_{t=1}^\mathbb{T}\left(r_{p,t} - \frac{1}{\mathbb{T}}\sum_{t=1}^{\mathbb{T}} r_{p,t}\right)^2}}, \quad \text{ESG}_{LQ} = \inf \left\{ x: \frac{1}{\mathbb{T}}\sum_{t=1}^{\mathbb{T}}\mathbb{I}(ESG_{p,t}\leq x)>0.25 \right\} \\ &
\text{ESG}_{M} = \frac{1}{\mathbb{T}}\sum_{t= 1}^{\mathbb{T}}ESG_{p,t}, \quad \text{SD}_{esg} = \sqrt{\frac{1}{\mathbb{T}-1}\sum_{t=1}^\mathbb{T}\left(ESG_{p,t} - \frac{1}{\mathbb{T}}\sum_{t=1}^{\mathbb{T}} ESG_{p,t}\right)^2},
\end{align*}
where $\mathbb{T}$ is the number of out-of-sample observations,  $r_{p,t} = \widehat{w}^\top r_{t}$ is the OOS portfolio excess return, $\widehat{w}$ is estimated portfolio and $r_{t}$ is the OOS excess assets return vector,  $ESG_{p,t} = \widehat{w}^\top ESG_t/\sum_{i=1}^p \widehat{w}_i$ is the standardized ESG score, $ESG_t$ is the OOS assets' ESG scores. 

In practical applications, transaction costs play a crucial role for investors. Following \cite{demiguel2009optimal, ao2019approaching} and \cite{fan2022}, the excess portfolio return net of transaction cost is computed as, 
\begin{equation}
r_{p}^{l,\text{net}} = \left(1 - \sum_{i=1}^pc\left|\widehat{w}_{l+1,i}-\widehat{w}_{l,i}^+\right| \right)(1+ r_{p}^l) -1,
	\label{eq transaction cost} 
\end{equation}
where $r_{p}^l$ is the excess return of the portfolio without transaction cost at the time of the $l$\textsuperscript{th} rebalancing, $\widehat{w}_{l+1}$ is the estimated portfolio weight at $(l+1)^{th}$ rebalancing, and $\widehat{w}_{l}^{+}$ represents the portfolio weight before the $(l+1)^{th}$ rebalancing.\footnote{The definition of $\widehat{w}_{l}^+$ can be found, e.g.,  \cite{demiguel2009optimal} and \cite{kan2022optimal}.} The parameter $c$ controls the level of the transaction cost, and in this study, it is set to 10 basis points following \cite{ao2019approaching}, \cite{ke2019predicting}, \cite{li2022synthetic}, and \cite{wu2025making}.  The Sharpe ratios considering transaction costs reported in the following tables are labeled as $\text{SR}_{tc}$.  Furthermore, we also report the portfolio turnover rate for each strategy, which is defined as $$\text{TO} = \sum_{l =1 }^{RT}\sum_{i=1}^p|\widehat{w}_{l+1,i}-\widehat{w}_{l,i}^+|/RT,$$ where $RT$ is the number of rebalancing.

\subsection{Results}

\begin{table}[h!t]
\centering
\caption{Empirical results when assets pool consists of largest 60 stocks in S\&P 500 index. Sample size is $90$. Out of sample period is from 06/2017 to 12/2024.}
\label{Table empirical largest 60}
\tabcolsep 0.156in
\begin{tabular}{lcccccc}
  \toprule             & SR    & SR$_{tc}$ & ESG$_M$ & ESG$_{LQ}$ & SD$_{esg}$ & TO      \\ \midrule
               & \multicolumn{6}{c}{$\bars = 0.9$}                               \\
Re-MV          & 0.243 & 0.234     & 0.915   & 0.899      & 0.024      & 0.435   \\
M-MV-S         & 0.174 & 0.127     & 0.936   & 0.803      & 0.178      & 592.475 \\
M-MV-POET      & 0.177 & 0.171     & 0.887   & 0.874      & 0.021      & 0.490   \\
M-MV-Linear    & 0.273 & 0.248     & 0.943   & 0.882      & 0.072      & 12.780  \\
M-MV-Nonlinear & 0.254 & 0.235     & 0.944   & 0.903      & 0.068      & 9.410   \\ \midrule
               & \multicolumn{6}{c}{$\bars = 0.8$}                               \\
Re-MV          & 0.240 & 0.233     & 0.825   & 0.807      & 0.021      & 0.446   \\
M-MV-S         & 0.193 & 0.134     & 0.841   & 0.724      & 0.150      & 303.992 \\
M-MV-POET      & 0.177 & 0.171     & 0.887   & 0.874      & 0.021      & 0.490   \\
M-MV-Linear    & 0.301 & 0.276     & 0.855   & 0.806      & 0.060      & 12.843  \\
M-MV-Nonlinear & 0.281 & 0.262     & 0.856   & 0.827      & 0.055      & 9.359   \\ \midrule
               & \multicolumn{6}{c}{Unconstrained Portfolio}                     \\
1/N            & 0.227 & 0.226     & 0.741   & 0.741      & 0.010      & 0.048   \\
Sample         & 0.205 & 0.154     & 0.693   & 0.583      & 0.135      & 257.467 \\
Q-MV           & 0.317 & 0.304     & 0.707   & 0.662      & 0.083      & 0.435   \\
MV-POET        & 0.216 & 0.212     & 0.766   & 0.756      & 0.022      & 0.393   \\
MV-Linear      & 0.343 & 0.318     & 0.697   & 0.654      & 0.085      & 13.315  \\
MV-Nonlinear   & 0.314 & 0.295     & 0.709   & 0.662      & 0.082      & 9.668   \\ \bottomrule
\end{tabular}
\end{table}

\begin{table}[h!t]
\centering
\caption{Empirical results when assets pool consists of largest 180 stocks in S\&P 500 index. Sample size is $90$. Out of sample period is from 06/2017 to 12/2024.}
\label{Table empirical largest 180}
\tabcolsep 0.156in
\begin{tabular}{lcccccc}
\toprule & SR     & SR$_{tc}$ & ESG$_M$ & ESG$_{LQ}$ & SD$_{esg}$ & TO      \\ \midrule
               & \multicolumn{6}{c}{$\bars = 0.9$}                                \\
Re-MV          & 0.197  & 0.188     & 0.903   & 0.890      & 0.020      & 1.151   \\
M-MV-S         & -      & -         & -       & -          & -          & -       \\
M-MV-POET      & -0.088 & -0.099    & 0.889   & 0.873      & 0.045      & 6.800   \\
M-MV-Linear    & 0.217  & 0.169     & 0.881   & 0.839      & 0.065      & 825.261 \\
M-MV-Nonlinear & 0.192  & 0.164     & 0.894   & 0.872      & 0.050      & 83.744  \\ \midrule
               & \multicolumn{6}{c}{$\bars = 0.8$}                                \\
Re-MV          & 0.216  & 0.208     & 0.819   & 0.806      & 0.017      & 1.164   \\
M-MV-S         & -      & -         & -       & -          & -          & -       \\
M-MV-POET      & -0.088 & -0.099    & 0.889   & 0.873      & 0.045      & 6.800   \\
M-MV-Linear    & 0.234  & 0.191     & 0.803   & 0.761      & 0.059      & 323.868 \\
M-MV-Nonlinear & 0.206  & 0.179     & 0.814   & 0.792      & 0.044      & 97.793  \\ \midrule
               & \multicolumn{6}{c}{Unconstrained Portfolio}                      \\
1/N            & 0.203  & 0.202     & 0.685   & 0.678      & 0.011      & 0.049   \\
Sample         & -      & -         & -       & -          & -          & -       \\
Q-MV           & 0.244  & 0.226     & 0.551   & 0.521      & 0.047      & 1.182   \\
MV-POET        & 0.191  & 0.173     & 0.680   & 0.671      & 0.028      & 1.350   \\
MV-Linear      & 0.278  & 0.239     & 0.544   & 0.504      & 0.061      & 149.910 \\
MV-Nonlinear   & 0.241  & 0.214     & 0.554   & 0.522      & 0.045      & 97.470  \\ \bottomrule
\end{tabular}
\end{table}

The empirical results are shown in Tables \ref{Table empirical largest 60}-\ref{Table empirical random 180}.  First, it can be observed that the newly proposed Re-MV portfolio, using $\widehat{\eta}^*\bQ$ that yields the highest estimated Sharpe ratio,  attains the top-ranked OOS SR, both with and without transaction costs, among all ESG-constrained portfolios. For example, in the case of the largest $p = 180$ stocks with target level $\bars = 0.9$ and  transaction costs, the SR of Re-MV is 0.188, which is the highest among all ESG-constrained portfolios. 
More importantly, compared with the naive M-MV-S strategy, the newly proposed  Re-MV outperforms it economically in all cases without sacrificing the ESG target, demonstrating the advantage of the new portfolio.  We note that, in some cases, M-MV-Li achieves high SR. However, this strategy is in fact unstable, as indicated by its high turnover rate. Our unreported results further show that its OOS returns are highly volatile.

\begin{table}[h!t]
\centering
\caption{Empirical results when assets pool consists of randomly selected 60 stocks in S\&P 500 index. Sample size is 90. Out of sample period is from 06/2017 to 12/2024.}
\label{Table empirical random 60}
\tabcolsep 0.156in
\begin{tabular}{lcccccc}
  \toprule             & SR     & SR$_{tc}$ & ESG$_M$ & ESG$_{LQ}$ & SD$_{esg}$ & TO      \\ \midrule
               & \multicolumn{6}{c}{$\bars = 0.9$}                                \\
Re-MV          & 0.206  & 0.196     & 0.899   & 0.870      & 0.044      & 0.339   \\
M-MV-S         & 0.071  & 0.026     & 0.799   & 0.647      & 0.193      & 349.939 \\
M-MV-POET      & -0.001 & -0.006    & 0.860   & 0.836      & 0.030      & 0.280   \\
M-MV-Linear    & 0.177  & 0.155     & 0.883   & 0.852      & 0.041      & 7.841   \\
M-MV-Nonlinear & 0.170  & 0.150     & 0.893   & 0.872      & 0.039      & 6.033   \\ \midrule
               & \multicolumn{6}{c}{$\bars = 0.8$}                                \\
Re-MV          & 0.218  & 0.209     & 0.823   & 0.797      & 0.032      & 0.339   \\
M-MV-S         & 0.080  & 0.031     & 0.749   & 0.622      & 0.175      & 263.183 \\
M-MV-POET      & -0.001 & -0.006    & 0.860   & 0.836      & 0.030      & 0.280   \\
M-MV-Linear    & 0.192  & 0.170     & 0.814   & 0.798      & 0.029      & 7.958   \\
M-MV-Nonlinear & 0.191  & 0.171     & 0.822   & 0.809      & 0.029      & 6.109   \\ \midrule
               & \multicolumn{6}{c}{Unconstrained Portfolio}                      \\
1/N            & 0.182  & 0.181     & 0.181   & 0.029      & 0.055      & 0.055   \\
Sample         & 0.084  & 0.036     & 0.560   & 0.188      & 359.352    & 359.352 \\
Q-MV           & 0.209  & 0.196     & 0.691   & 0.027      & 0.337      & 0.337   \\
MV-POET        & 0.054  & 0.050     & 0.627   & 0.028      & 0.192      & 0.192   \\
MV-Linear      & 0.205  & 0.184     & 0.688   & 0.026      & 8.056      & 8.056   \\
MV-Nonlinear   & 0.209  & 0.190     & 0.690   & 0.028      & 6.197      & 6.197  \\ \bottomrule
\end{tabular}
\end{table}

\begin{table}[h!t]
\centering
\caption{Empirical results when assets pool consists of randomly selected 180 stocks in S\&P 500 index. Sample size is 90. Out of sample period is from 06/2017 to 12/2024.}
\label{Table empirical random 180}
\tabcolsep 0.156in
\begin{tabular}{lcccccc}
     \toprule          & SR    & SR$_{tc}$ & ESG$_M$ & ESG$_{LQ}$ & SD$_{esg}$ & TO      \\ \midrule
               & \multicolumn{6}{c}{$\bars = 0.9$}                               \\
Re-MV          & 0.185 & 0.176     & 0.895   & 0.890      & 0.037      & 1.018   \\
M-MV-S         & -     & -         & -       & -          & -          & -       \\
M-MV-POET      & 0.170 & 0.154     & 0.892   & 0.860      & 0.040      & 0.802   \\
M-MV-Linear    & 0.187 & 0.160     & 0.860   & 0.832      & 0.051      & 68.534  \\
M-MV-Nonlinear & 0.184 & 0.159     & 0.815   & 0.770      & 0.082      & 45.729  \\ \midrule
               & \multicolumn{6}{c}{$\bars = 0.8$}                               \\
Re-MV          & 0.209 & 0.201     & 0.793   & 0.773      & 0.028      & 1.017   \\
M-MV-S         & -     & -         & -       & -          & -          & -       \\
M-MV-POET      & 0.170 & 0.154     & 0.892   & 0.860      & 0.040      & 0.802   \\
M-MV-Linear    & 0.198 & 0.174     & 0.793   & 0.770      & 0.036      & 85.604  \\
M-MV-Nonlinear & 0.199 & 0.174     & 0.749   & 0.719      & 0.065      & 44.631  \\ \midrule
               & \multicolumn{6}{c}{Unconstrained Portfolio}                     \\
1/N            & 0.203 & 0.202     & 0.202   & 0.202      & 0.022      & 0.058   \\
Sample         & -     & -         & -       & -          & -          & -       \\
Q-MV           & 0.236 & 0.221     & 0.529   & 0.448      & 0.091      & 1.013   \\
MV-POET        & 0.177 & 0.166     & 0.641   & 0.620      & 0.027      & 0.594   \\
MV-Linear      & 0.231 & 0.209     & 0.613   & 0.597      & 0.036      & 123.699 \\
MV-Nonlinear   & 0.239 & 0.212     & 0.534   & 0.458      & 0.089      & 41.629 \\ \bottomrule
\end{tabular}
\end{table}

Second, in terms of ESG measures, all ESG-constrained strategies achieve OOS ESG scores close to the prespecified threshold, and the proposed portfolio satisfies this threshold best overall. For example, in the case of the largest $p = 180$ stocks with target level  $\bars = 0.9$,  the average OOS ESG score of Re-MV is 0.903, while the second-closest (to target) ESG score is achieved by M-MV-Nonlinear at 0.894. Furthermore, the OOS ESG scores of Re-MV exhibit low variability. On the one hand, the lower 25$^{th}$ percentile of the OOS ESG scores remains close to the target level, indicating that the OOS portfolio ESG scores stay near the target in most periods; In contrast, the lower 25$^{th}$ percentile of ESG score for other constrained portfolios (particularly M-MV-Linear) shows a substantial larger deviation from the target level.
On the other hand, the  standard deviation of ESG score of Re-MV is either the lowest or very close to the lowest among all competing strategies.
As a result, our regularized ESG-constrained portfolio  stably achieves a sustainable ESG level while maintaining a high Sharpe ratio.  Additionally, we can observe that the proposed Re-MV  exhibits a low turnover rate. For example, in the case of the  largest $p = 60$ stocks, the Re-MV strategy exhibits a turnover rate of 1.151, which is the second lowest among the considered strategies. The lowest turnover rate is achieved by the equal weighting strategy. As a result, when transaction costs are taken into account (or play more important role), the superiority of the Re-MV strategy becomes more pronounced.  Lastly, compared with the corresponding unconstrained strategies, imposing an ESG-level constraint generally reduces Sharpe ratio performance while improving ESG performance. A tighter constraint (with $\bars$ increasing from 0.8 to 0.9) leads to a larger decrease in OOS SR. However, it is worth noting that, compared to Q-MV, the proposed Re-MV can possibly achieve a higher Sharpe ratio, as indicated by the results in Lemma \ref{lemma OOS SR asymtotic} and Remark \ref{remark unconstrained port}. This theoretical insight is further supported by our empirical results with randomly selected assets. As reported in Table \ref{Table empirical random 60}, when  $\bars = 0.8$ without transaction costs, the SR of Re-MV is 0.218, which exceeds the corresponding value of 0.209 for Q-MV.

\section{Conclusion}
\label{sec: conclude}
This paper studies a regularized ESG-constrained mean-variance (MV) portfolio optimization in large-dimensional situation.  We first quantify the impact of imposing an ESG-level constraint, as well as the estimation error arising from the sample mean estimation of the ESG score mean level, by deriving the limiting OOS Sharpe ratio of the proposed portfolio. Then, we further propose an estimator for OOS SR such that we can evaluate the impacts of various specification of regularization matrix based on historical data. The theoretical consistency is also provided.

Simulation results show that the proposed estimators perform close to the corresponding unknown oracle counterparts. Moreover, we numerically investigate the impact of various forms of regularization matrices on the OOS SR, which provides useful guidance for practical implementation.
Finally, based on OOS SR estimator, we propose an adaptive regularized portfolio which uses the best regularization matrix yielding the highest estimated SR (among a set of candidates) at each decision node.
Empirical evidence based on the S\&P 500 index demonstrates that the proposed adaptive ESG-constrained portfolio achieves a high OOS SR while satisfying the required ESG level. Overall, this study provides a statistically rigorous and practically effective solution for integrating sustainability into high-dimensional portfolio allocation.

\bibliography{reference}

\end{document}